\documentclass[10pt]{iopart}
\usepackage[dvips]{graphicx}
\usepackage[hypertex]{hyperref}
\usepackage{iopams}

\usepackage{bm}

\newcommand{\vk}{\mathbf}

\newcommand{\ZETF}{Zh. Eksp. Teor. Fiz. }

\newcommand{\m}{\mathrm}
\newcommand{\tinv} {{\cal{T}}}

\newcommand{\prt} {\partial}
\newcommand{\up} {u_{p}}

\newcommand{\vp}{{\bf p}}
\newcommand{\vP}{{\bf P}}
\newcommand{\vr}{{\bf r}}

\newcommand{\vF}{{\bf F}}

\newcommand{\vq}{{\bf q}}
\newcommand{\vQ}{{\bf Q}}
\newcommand{\ve}{{\bf e}}
\newcommand{\vd}{{\bf d}}
\newcommand{\sg}{\hat{s}}

\newcommand{\dir} {{(\m{dr})}}
\newcommand{\resc} {{(\m{rsc})}}
\newcommand{\rescI} {{(\m{rsc}1)}}
\newcommand{\rescII} {{(\m{rsc}2)}}

\newcommand{\ov} {\overline}
\newcommand{\tl}{\tilde}

\newcommand{\tld} {{\tl{\vk{d}}_n}}
\newcommand{\hatd} {{\hat{\vk{d}}}_n}

\newcommand{\mcal} {\mathcal}

\begin{document}

\title[Rescattering effects in
laser-assisted electron-atom bremsstrahlung]
{Rescattering effects in
laser-assisted electron-atom bremsstrahlung}

\author{A~N~Zheltukhin$^1$, A~V~Flegel$^2$, M~V~Frolov$^1$, N~L~Manakov$^1$ and Anthony~F~Starace$^3$}

\address{$^1$Department of Physics,  Voronezh State University, Voronezh 394006, Russia}
\address{$^2$Department of Computer Science,  Voronezh State University,
Voronezh 394006, Russia}
\address{$^3$Department of Physics and Astronomy, University of Nebraska, Lincoln, Nebraska 68588-0299, USA}

\begin{abstract}

Rescattering effects in nonresonant spontaneous laser-assisted
electron-atom bremsstrahlung (LABrS) are analyzed within the framework of time-dependent effective-range (TDER) theory. It is shown that high energy LABrS spectra exhibit rescattering plateau structures that are similar to those that are
well-known in strong field laser-induced processes as well as those that have been predicted theoretically in laser-assisted collision processes. In the limit of a low-frequency laser field, an analytic description
of LABrS  is obtained from a rigorous quantum
analysis of the exact TDER results for the LABrS amplitude. This
amplitude is represented as a sum of factorized terms involving three
factors, each having a clear physical meaning.  The first two factors are the exact field-free
amplitudes for electron-atom bremsstrahlung and for electron-atom
scattering, and the third factor describes free electron motion in the
laser field along a closed trajectory between the first
(scattering) and second (rescattering) collision events. Finally, a
generalization of these TDER results to the case of
LABrS in a Coulomb field is discussed.

\end{abstract}


\maketitle

\ioptwocol

\section{Introduction}

An intense laser field significantly modifies the bremsstrahlung (BrS) process accompanying
electron-atom or electron-ion scattering, i.e., an electron colliding with a target can efficiently
convert the combined energies of a number of laser photons, each with energy $\hbar\omega$, into the energy of a
spontaneously emitted photon $\hbar\Omega$. In comparison with field-free BrS, in laser-assisted
BrS (LABrS) the spectral energies can be significantly extended and resonant-like enhancements of
the LABrS cross sections may appear.

There have been relatively few prior studies of LABrS processes. Karapetyan and Fedorov~\cite{KF}
have shown that LABrS spectra  for electron scattering from a Coulomb potential exhibit resonant
peaks (at $\Omega = k\omega$, where $k$ is integer), which occur only in the limit of an intense
laser field. The electron-Coulomb interaction was treated in~\cite{KF} within the Born
approximation. Zhou and Rosenberg~\cite{Rs} investigated LABrS beyond the Born
approximation in the scattering potential for the case of a low-frequency laser field.  They found
a series of resonant peaks in the BrS spectrum, separated by $\hbar\omega$, that are related to
resonant features in field-free electron-atom scattering. The approach used in~\cite{Rs} is based
on the low-frequency Kroll-Watson result~\cite{KW} for the electron scattering state and its
``resonant'' modification. These two seminal works by Karapetyan and Fedorov~\cite{KF} and by Zhou
and Rosenberg~\cite{Rs} have led to a number of more recent analyses. These have been described
briefly in section 4.5 of the review of Ehlotzky \textit{et al.}~\cite{Ehlotzky1998}.  A
comparative analysis of LABrS for electron-Coulomb scattering within the Born and low-frequency
approximations (in accordance with~\cite{KF} and~\cite{Rs} respectively) has been given in a recent
paper by Dondera and Florescu~\cite{obs}, who also review there other works on non-relativistic
LABrS.

A common feature of the above works is that they do not describe effects of laser-induced
electron rescattering on an atomic target. These rescattering effects are well known in multiphoton
processes involving bound atomic states, such as high-order harmonic generation (HHG) and
above-threshold ionization/detachment (ATI/ATD). In those processes they lead to the appearance of broad, plateau-like
structures in the HHG and ATI/ATD spectra~\cite{Salieres, Becker02, Ehlotzky1}. Also, plateaus in
the high-energy (multiphoton) spectra of laser-assisted collisional processes involving an initially free (i.e.,
continuum) electron have been predicted for laser-assisted electron scattering
(LAES)~\cite{jetpl02, Milo04} and laser-assisted radiative recombination/attachment
(LARR/LARA)~\cite{MiloEhl02, eff_at_p}. Interest in such plateaus centers on the possibility of
transferring large amounts of energy from a laser field into either electron kinetic energy or
high-energy spontaneous photons (or harmonics) without significant decreases in the yields as the number
$n$ of absorbed photons increases over a wide interval of $n$.

The theoretical description of laser-assisted electron-atom processes for the case of an intense
laser field necessarily requires an accurate treatment of both the electron-laser and electron-atom
interactions in order to properly describe the strong field plateau phenomena. An appropriate
approach that meets this requirement is based on time-dependent effective range (TDER)
theory~\cite{TDER03}, which combines the effective range theory (for the electron-atom interaction)
with the quasienergy or Floquet theory (for the electron-laser field interaction). This approach
was employed recently to describe resonant phenomena in the LABrS process~\cite{ResBrs}. Namely, a
resonant mechanism for LABrS involving the resonant transition into a laser-dressed intermediate
quasi-bound state (corresponding respectively to  a field-free bound state of either a neutral atom
or a negative ion) accompanied by ionization or detachment of this state by the laser field has been
considered. However, rescattering effects were not investigated in~\cite{ResBrs}. An important
advantage of the TDER theory is that it allows an accurate quantum derivation of closed-form
analytic formulas for the cross sections of strong field processes that take into account the rescattering
effects non-perturbatively in the limit of a low frequency laser field. Such formulas were obtained
for both laser-induced processes ( such as HHG~\cite{JPBHHG2009} and ATI/ATD~\cite{analit_atd}) and
laser-assisted collisional processes (such as LAES~\cite{analit_laes, LAES2013} and
LARR/LARA~\cite{eff_at_p}). For collisional processes, the analytic formulas describe accurately the
high-energy part of the electron (for LAES) or photon (for LARR/LARA) spectra, which cannot be
described using the well-known Bunkin-Fedorov~\cite{BF} or Kroll-Watson~\cite{KW} approximations.
The analytic results have a factorized structure, in which the atomic factors represent exact field-free
amplitudes evaluated using the instantaneous kinetic electron momenta in the laser field, thus allowing us to generalize
the effective range approximation to the case of an arbitrary atomic potential.

In this paper, we consider rescattering effects in nonresonant LABrS, i.e., we study the
process of spontaneous photon emission during electron-atom scattering for electron energies and
laser field parameters such that resonant radiative (spontaneous or stimulated) transitions into an
intermediate quasi-bound state are negligible and thus are not taken into account.
In the low-frequency approximation, we derive an analytic description of the rescattering plateau
features in LABrS spectra and analyze numerically the accuracy of the derived analytic formulas.
These analytic results permit a transparent physical interpretation and present the first quantum justification of the three-step rescattering scenario for the description of
radiative (with emission of a spontaneous photon) continuum-continuum transitions of an electron
interacting with both an atomic potential and an intense laser field.

This paper is organized as follows. In section 2 we present results for the LABrS
amplitude in terms of the exact TDER expressions for initial and final states of the scattered
electron. In section 3 we present an analysis of the LABrS amplitude in the limit of low frequencies. We start in section 3.1 with the low-frequency expansion of the most important ingredient determining a TDER scattering state. In section 3.2 we present the low-frequency result for the amplitude for
``direct'' LABrS, i.e., neglecting rescattering. Basic results (\ref{d:resc2}) -- (\ref{tdp2}) for the
``rescattering'' part, $\vk{d}^{\resc}_n$, of the LABrS amplitude are obtained in section 3.3 using
the regular saddle point method. Based on results of section 3.3, in section 3.4 we discuss the
three-step rescattering scenario for the LABrS process and its relation to the corresponding scenario for LAES. In section 3.5 we consider the case of two merging saddle points, which requires the use
of special saddle point methods in order to estimate the contributions of coalescing electron
trajectories to the rescattering amplitude $\vk{d}^{\resc}_n$. Our numerical results and
discussions are presented in section 4. In section 4.1 we demonstrate good agreement between exact
TDER results and those obtained in the low-frequency approximation for the high-energy rescattering plateau part of LABrS spectra. We discuss there also some peculiarities in LABrS spectra (such as
their oscillation patterns and interference enhancements). In section 4.2 we generalize
our TDER results, which are valid for a short-range atomic potential, to the case of LABrS in a Coulomb potential and
present numerical estimates for electron-proton LABrS. In section 5 we present our conclusions.
In the Appendix we present some mathematical details.

\section{General results for LABrS within TDER theory}\label{GenResults}

We consider the LABrS process for the case of a linearly polarized monochromatic laser field described by the
electric field vector $\vF(t)$,
\begin{equation}
\label{Ft}
 \vk{F}(t)=\vk{e}_z F\cos\omega t,
\end{equation}
where $F$ is  the field amplitude and $\vk{e}_z$ is the unit polarization vector. For the
electron-laser interaction, we use the dipole approximation in the length gauge
$V(\vk{r},t)=-e\vk{r}\cdot\vk{F}(t)$. To describe electron-atom collisions in a time-periodic
field~(\ref{Ft}), the quasienergy (or Floquet) approach is most appropriate. Within this approach,
the laser-dressed scattering state of an electron with asymptotic momentum $\vk{p}$ and kinetic
energy $E=p^2/(2m)$ has the form (cf., e.g.,~\cite{PhysRep86}):
\begin{eqnarray}
\nonumber &&\Psi_{\vk{p}}(\vk{r},t)=\Phi_{\vk{p}}(\vk{r},t)e^{-i\epsilon
t/\hbar}, \\
&&\Phi_{\vk{p}}(\vk{r},t)=\Phi_{\vk{p}}(\vk{r},t+T), \quad T=2\pi/\omega, \label{QES}
\end{eqnarray}
where $\epsilon$ is the quasienergy, $ \epsilon = E+\up$, and $\up=e^2 F^2/(4m\omega^2)$ is the
ponderomotive (or quiver) energy of an electron in the laser field~(\ref{Ft}).

For given initial ($\vk{p}_{i}$) and final ($\vk{p}_{f}$) electron momenta, the LABrS process
consists in the spontaneous emission of a photon with energy
$\hbar\Omega=(p_i^2-p_f^2)/(2m)+n\hbar\omega$ (where $n$ is the number of absorbed, $n>0$, or
emitted, $n<0$, laser photons) and polarization vector $\ve'$ $(\ve'\cdot \ve'^*=1)$. The amplitude
for this process is proportional to the scalar product $\vk{e}'^*\cdot \vk{d}_n$ of the polarization
vector and the Fourier component $\vk{d}_n\equiv \vk{d}_n(\vk{p}_{i},\vk{p}_{f})$ of the dipole matrix
element $\langle \Phi_{\vk{p}_{f}}^{\tinv}|\vk{d}|\Phi_{\vk{p}_{i}}^{}\rangle$,
\begin{equation}
\label{dip}
\vk{d}_n=\frac{1}{T}\int\limits_{0}^{T}dt e^{i n\omega t } \langle
\Phi_{\vk{p}_{f}}^{\tinv}(t)|\vk{d}|\Phi_{\vk{p}_{i}}^{}(t)\rangle, \quad \vk{d} = e\vk{r},
\end{equation}
where $\Phi_{\vk{p}}(t)\equiv \Phi_{\vk{p}}(\vk{r},t)$ and $\Phi_{\vk{p}}^{\tinv}(\vk{r},t)$
[$=\Phi_{-\vk{p}}^{*}(\vk{r},-t)$] is the corresponding time-reversed wave function~\cite{ResBrs}. The LABrS doubly
differential cross section with respect to the emitted photon frequency $\Omega$ and the final
electron direction (into the solid angle element $d\Omega_{\vp_f}$), summed
over polarizations and integrated over the directions of the emitted photon, has the following form:
\begin{equation}\label{CS}
\frac{ d^2\sigma_n(\vk{p}_{i},\vk{p}_f)}{d\Omega d\Omega_{\vk{p}_f}}= \frac{m^2\Omega^3 }{6 (\pi
\hbar
c)^3}\frac{p_{f}}{p_{i}}|\vk{d}_n(\vk{p}_{i},\vk{p}_{f})|^2.
\end{equation}

We use the TDER theory to describe the field-dressed continuum state ($\Phi_{\vp}$) of the active
electron~\cite{LAES2013,jetpl08} scattered from a short-range atomic potential $U(r)$ that vanishes
for $r\gtrsim r_c$. The TDER theory assumes that the potential $U(r)$ supports a single
weakly-bound state (a negative ion state) with energy $E_0=-\hbar^2\kappa^2/(2m)$ $(\kappa r_c \ll
1)$ and angular momentum $l$. The electron-atom interaction is described by the $l$-wave scattering
phase $\delta_l(p)$, which is parameterized by the scattering length $a_l$ and the effective range
$r_l$, which are parameters of the problem:
\begin{equation*}
k^{2l+1}\cot\delta_l(p)=-a_l^{-1}+r_lk^2/2,\quad k=p/\hbar.
\end{equation*}
For simplicity, in this paper we consider the case of a bound $s$-state ($l=0$), so that only the
phase shift $\delta_0(p)$ is nonzero.

The scattering state $\Phi_{\vp}(\vr,t)$ may be presented as a sum of the ``incident'' plane wave,
$\chi_{\vk{p}}(\vk{r},t)$, and the scattered ``outgoing'' wave, $\Phi_{\vk{p}}^{(+)}(\vk{r},t)$,
\begin{equation}
\label{scatt.state} \Phi_{\vk{p}}(\vk{r},t) =
\chi_{\vk{p}}(\vk{r},t)+\Phi_{\vk{p}}^{(+)}(\vk{r},t).
\end{equation}
The function  $\chi_{\vk{p}}$ is the time-periodic part of the wave function of a free electron with
momentum $\vk{p}$ in the laser field $\vF(t)$ (the Volkov wave function),
\begin{eqnarray}
&& \chi_{\vk{p}}(\vk{r},t) = e^{i
[\vk{P}(t)\cdot\vk{r}+S_{\vk{p}}(t)]/\hbar},\nonumber\\
&&S_{\vk{p}}(t)=-\int^t [\vk{P}^2(\tau)/(2m)-\epsilon]d\tau \nonumber \\
&&= \frac{e F \vk{e}_z\cdot \vk{p}}{m\omega^2}\cos\omega t + \frac{u_p}{2\omega}\sin 2\omega t,
\label{vac}
\end{eqnarray}
where $\vk{P}(t)=\vk{p}-(e/c)\vk{A}(t)$ is the electron's kinetic momentum in the laser field,
while $\vk{A}(t)=-(\vk{e}_z  F c /\omega)\sin\omega t$ is the vector potential of the laser field.

Within the TDER approach, the scattered electron wave function $\Phi_{\vk{p}}^{(+)}$ in
(\ref{scatt.state}) is expressed in terms of a one-dimensional integral~\cite{jetpl08}, involving
the retarded Volkov Green's function, $G^{(+)}(\vr,t;\vr',t')$, for a free electron in the laser
field $\vF(t)$:
\begin{eqnarray}
\nonumber \Phi_{\vk{p}}^{(+)}(\vk{r},t) &=& -\frac{2\pi\hbar^2}{m}\int_{-\infty}^t dt'
e^{i \epsilon (t-t')/\hbar} \\
&\times& G^{(+)}(\vk{r},t;0,t')f_{\vk{p}}(t'),\label{Phi+}
\end{eqnarray}
The time-inverted scattering state $\Phi_{\vk{p}}^{\tinv}$, involved in the transition matrix
element in (\ref{dip}), has the form~(\ref{scatt.state}) with the scattered wave $\Phi_{\vp}^{(+)}$
replaced by the wave function $\Phi_{\vp}^{(-)}$: $\Phi_{\vk{p}}^{\tinv} =
\chi_{\vp}+\Phi_{\vp}^{(-)}$. The function $\Phi_{\vp}^{(-)}$ has the asymptotic form of an ``ingoing''
wave and is expressed in terms of the advanced Volkov Green's function, $G^{(-)}$:
\begin{eqnarray}
\nonumber \Phi_{\vk{p}}^{(-)}(\vk{r},t) &=& -\frac{2\pi\hbar^2}{m}\int_{t}^{\infty} dt'
e^{i \epsilon (t-t')/\hbar} \\
&\times& G^{(-)}(\vk{r},t;0,t')f^{*}_{-\vk{p}}(-t'). \label{Phi-}
\end{eqnarray}
For the Volkov Green's functions, $G^{(\pm)}$, we use the well-known Feynman form:
\begin{eqnarray}
\nonumber
G^{(\pm)}(\vr,t;\vr',t') &=& \mp\theta[\pm(t-t')]\frac{i}{\hbar}\left[ \frac{m}{2\pi i\hbar(t-t')} \right]^{3/2}\\
&\times & \exp[iS(\vr,t;\vr',t')/\hbar], \label{Green}
\end{eqnarray}
where $S(\vr,t;\vr',t')$ is the classical action of the active electron in the field $\vk{F}(t)$ (cf.,
e.g.,~\cite{ResBrs}). In the absence of the laser field ($\vF(t)=0$), the functions
$\Phi_{\vp}^{(\pm)}$ reduce to the wave functions $\psi_\vp^{(\pm)}$ for the field-free problem of
elastic electron scattering on the potential $U(r)$ having asymptotic forms of spherical outgoing (+) or
ingoing (-) waves:
\[
\psi_\vp^{(\pm)}(r)\big|_{r\gg r_c} =  \mathcal{A}(p)\frac{e^{\pm
i p r/\hbar}}{r},
\]
where $\mathcal{A}(p)$ is the field-free $s$-wave scattering amplitude in the effective range
approximation:
\begin{equation}
\label{scat.ampl} \mathcal{A}(p)=\frac{1}{-a_0^{-1}-i p/\hbar + r_0p^2/(2\hbar^2).}
\end{equation}

The time-periodic function $f_\vp(t)=\sum_k f_k(\vp)e^{-ik\omega t}$ appearing in the wave
functions~(\ref{Phi+}) and~(\ref{Phi-}) is the key object of TDER theory. It contains the
entire information about the details of the electron-atom dynamics in the laser field and enters into
the expressions for the amplitudes of laser-assisted collisions (cf., e.g., the results in~\cite{jetpl08}
for the LAES amplitude). The function $f_{\vp}(t)$ satisfies an inhomogeneous integro-differential
equation (cf.~equation~(23) in \cite{LAES2013}), which can be converted to a system of
inhomogeneous linear equations for its Fourier coefficients $f_k(\vp)$ (cf. Appendix~A in
\cite{LAES2013}).

The TDER result for the LABrS dipole moment $\vk{d}_n$ [given by expression~(\ref{dip})] follows
after substituting the wave functions of the initial ($\Phi_{\vp_i}$) and final ($\Phi_{\vp_f}$)
scattering states in the form~(\ref{scatt.state}) [taking into account (\ref{Phi+})
and~(\ref{Phi-})] into equation (\ref{dip}) (for details of the derivation, see~\cite{ResBrs}):
\begin{equation}\label{dip111}
\vk{d}_n =\vk{d}^{(TS)}_n + \hatd + \tld.
\end{equation}
In the result~(\ref{dip111}), the term $\vk{d}^{(TS)}_n$ corresponds to the Thomson scattering (TS)
of the laser radiation from a free electron,
\[
\vk{d}_n^{(TS)}=\frac{1}{T}\int\limits_{0}^{T}dt\, e^{i n\omega t }
\langle\chi_{\vk{p}_{f}}(t)|\vk{d}|\chi_{\vk{p}_{i}}(t)\rangle,
\]
and is nonzero only for $\Omega=\omega$ $(n=1)$. In what follows, we omit this term from our
considerations. Other terms in expression (\ref{dip111}) are given by the following matrix
elements:
\begin{eqnarray}
\nonumber && \hatd = \frac{1}{T}\int\limits_{0}^{T}dt e^{i
n\omega t } \Big(\langle\chi_{\vk{p}_{f}}(t)|\vk{d}|\Phi_{\vk{p}_{i}}^{(+)}(t)\rangle\\
&& \phantom{{\hatd}} + \langle\Phi_{\vk{p}_{f}}^{(-)}(t)|\vk{d}|\chi_{\vk{p}_{i}}^{}(t)\rangle
\Big), \label{hatdip}\\
&& \label{Scatt-Scatt} \tld = \frac{1}{T}\int\limits_{0}^{T}dt e^{i n\omega t }
\langle\Phi_{\vk{p}_{f}}^{(-)}(t)|\vk{d}|\Phi_{\vk{p}_{i}}^{(+)}(t)\rangle.
\end{eqnarray}

The derivation of the final results for the dipole matrix elements (\ref{hatdip}) and~(\ref{Scatt-Scatt}) without using any
approximations for the TDER scattering states $\Phi_{\vk{p}_{i}}^{(+)}$ and
$\Phi_{\vk{p}_{f}}^{(-)}$ can be found in~\cite{ResBrs}. These
exact TDER results for the LABrS amplitude are complicated since they contain infinite series involving Fourier
coefficients of the functions $f_{\vk{p}_i}(t)$ and $f_{-\vk{p}_f}(-t)$, which can be obtained only
numerically. For a non-perturbative analytic analysis of rescattering effects,
we shall thus employ the low-frequency approximation.

\section{Low-frequency analysis of the LABrS amplitude}

\subsection{Low-frequency results for $f_{\vp}(t)$}

In a low-frequency laser field (such that the condition $\hbar\omega \ll u_p$ is fulfilled), the
equation for $f_\vp(t)$ can be solved analytically using an iterative approach for taking into
account the rescattering effects. This iterative approach has been developed in~\cite{LAES2013}.
As a result, the function $f_\vp(t)$ can be presented as a sum of two terms,
\begin{equation}
\label{f(t):sum}
f_{\vk{p}}(t) = f^{\dir}_{\vk{p}}(t)+f^{\resc}_{\vk{p}}(t),
\end{equation}
where the first term corresponds to the Kroll-Watson
approximation for the scattering state~\cite{analit_laes, LAES2013}:
\begin{eqnarray}
\label{f(t):dr} f^{\dir}_{\vk{p}}(t)=
\mathcal{A}[P(t)]e^{iS_{\vk{p}}(t)/\hbar},
\end{eqnarray}
which describes the ``direct'' (without rescattering) LAES. The second term
in~(\ref{f(t):sum}) represents the first-order ``rescattering'' correction to the zero-order
result~(\ref{f(t):dr}). It involves the product of two field-free electron-atom
scattering amplitudes with laser-modified instantaneous momenta, thus describing
electron-atom rescattering (cf.~\cite{LAES2013}):
\begin{eqnarray}
\label{f(t):rsc}
 &&f^{\resc}_{\vk{p}}(t)=  \sum_{k}  \mcal{A}\left[Q(t,t'_k)\right] \chi_\vk{p}(t,t'_k),\\
 &&\chi_{\vk{p}}(t,t'_k)=\frac{
\mcal{A}\left[P(t'_k)\right]
e^{i\varphi_{\vk{p}}(t,t'_k)}}{\sqrt{(\hbar/m)(t-t'_k)^3
 \mcal{D}_{\vk{p}}(t,t'_k)}},\label{chi_p}
\end{eqnarray}
where
\begin{eqnarray}
\label{Q} && \vk{Q}(t,t')=\vk{q}(t,t')-\frac{e}{c}\vk{A}(t),
\\
\label{q} && \vk{q}(t,t')
=\frac{e}{\omega^2}\frac{\vk{F}(t)-\vk{F}(t')}{t-t'},
\\
 &&
\varphi_{\vk{p}}(t,t')=\left[\epsilon(t-t')+S(t,t')+S_{\vk{p}}(t')\right]/\hbar
\nonumber\\
&&\label{phi_def}
=\frac{(t-t')(\vk{p}-\vk{q}(t,t'))^2}{2m\hbar}+\frac{S_{\vk{p}}(t)}{\hbar},
\\ &&
\nonumber \mcal{D}_{\vk{p}}(t,t')= \prt^2
\varphi_{\vk{p}}(t,t')/\prt t'^2
\\
&&=\frac{ \vk{Q}^2(t',t)}{m \hbar(t-t')}+\frac{e
\vk{F}(t')\cdot(\vk{q}(t,t')-\vk{p})}{m \hbar},\label{D_def}
\end{eqnarray}
where  $S(t,t')\equiv  S(0,t;0,t')$. For $\vk{Q}(t,t')$
in~(\ref{Q}) we have the following relation:
\[
\vk{Q}(t',t) = \vk{Q}(t,t') + \frac{e}{c}\vk{A}(t) -
\frac{e}{c}\vk{A}(t').
\]
The summation in~(\ref{f(t):rsc}) is taken over the saddle points
$t'_k$ of the phase function $\varphi_\vp(t,t')$ given by
equation~(\ref{phi_def}). These saddle points satisfy the equation
$(\prt\varphi_{\vk{p}}(t,t')/\prt t')|_{t'=t'_k}=0$ or,
explicitly,
\begin{equation}
\label{sp1} \vk{P}^2(t'_k) = \vk{Q}^2(t'_k,t).
\end{equation}
The equation~(\ref{sp1}) has a clear physical meaning: it represents the energy conservation law for
the electron scattering event at the time $t'_k$ with exchange of the kinetic momentum from $\vk{P}(t'_k)$
to the ``intermediate'' laser-induced momentum $\vk{Q}(t'_k,t)$. This latter momentum ensures the
condition for the electron return by the laser field back to the atom at the time moment $t$,
followed by the recollision. We note that the saddle points $t_k'$ are functions of the time $t$
($t_k'\equiv t_k'(t)$) implicitly defined by equation~(\ref{sp1}).

The results~(\ref{f(t):dr}) and~(\ref{f(t):rsc}) are not
applicable for resonant electron energies,
$E\approx\mu\hbar\omega+E_0-u_p$, at which the electron may be
temporarily captured in a bound state of the atomic potential by
emitting $\mu$ photons~\cite{res_laes}. Nor are they applicable
for threshold energies, $E=\nu \hbar\omega,$ $\nu =1,2,\ldots$, at
which the function $f_{\vk{p}}(t)$ may be affected considerably by
threshold phenomena, corresponding to the closing (or opening) of
the channel for stimulated emission of $\nu$ laser photons by the
incident electron~\cite{jetpl08}. Thus, in this paper we consider
the LABrS process only for non-resonant and non-threshold
conditions.

\subsection{Low-frequency result for the ``direct'' LABrS amplitude $\vk{d}^{\dir}_n$  }

Using the ``rescattering'' expansion~(\ref{f(t):sum}) for the function $f_\vp(t)$, which determines the wave functions
$\Phi_{\vk{p}_{i}}^{(+)}(\vr,t)$ and $\Phi_{\vk{p}_{f}}^{(-)}(\vr,t)$, we can build the ``rescattering'' expansion for
the LABrS dipole moment $\vk{d}_n$, representing it in the form:
\begin{equation}
\label{dip:exp}
\vk{d}_n = \vk{d}^{\dir}_n +\vk{d}^{\resc}_n,
\end{equation}
where the term $\vk{d}^{\dir}_n$ corresponds to the ``direct'' LABrS process, while the term
$\vk{d}^{\resc}_n$ (the first order ``rescattering'' correction to $\vk{d}^{\dir}_n$) describes the
rescattering effects in the LABrS amplitude. The ``direct'' term $\vk{d}^{\dir}_n$
may be obtained from the dipole moment $\hatd$ given by expression~(\ref{hatdip}). The matrix
element in equation~(\ref{Scatt-Scatt}) for $\tld$ contains the product of two functions,
$f_{\vp_i}(t)$ and $f_{-\vp_f}(-t)$ [cf.~(\ref{Phi+}) and~(\ref{Phi-})], each of them being
proportional to the field-free electron-atom scattering amplitude, which means that $\tld$ describes
rescattering effects in LABrS and should thus be neglected when calculating the zero-order term
$\vk{d}^{\dir}_n$.

Let us analyze first the dipole moment $\hatd$. It has the following explicit form
(cf.~\cite{ResBrs}):
\begin{eqnarray}
\label{dipole:chi12}\nonumber &&\hatd =\frac{\pi\hbar e^2F}{
m^2\Omega^2\omega} \Bigg[ \frac{2i\omega}{eF}\left(
\vk{p}_i\mathcal{L}^{(1)}_n - \vk{p}_f\mathcal{L}^{(2)}_{n}\right)\\
\label{dipole:chi3} \nonumber && +\vk{e}_z
\sum_{s=\pm1}\frac{\mathcal{L}^{(1)}_{n+s} - \mathcal{L}^{(2)}_{n+s}}
{\omega/\Omega+s}\Bigg],
\end{eqnarray}
where
\begin{eqnarray}
\label{L12:int}
&& \mathcal{L}_{n}^{(1,2)} = \frac{1}{T}\int_0^T dt e^{in\omega t}L^{(1,2)}(t),
\end{eqnarray}
where
\begin{eqnarray}
\label{L1}
&& L^{(1)}(t)=f_{-\vp_f}(-t) e^{iS_{\vp_i}(t)/\hbar}, \\
\label{L2}
&& L^{(2)}(t)=f_{\vp_i}(t) e^{iS_{-\vp_f}(-t)/\hbar}.
\end{eqnarray}
We restrict our considerations to the high-energy LABrS spectrum, in which the emitted
photon energy is much greater than the laser photon energy,
\begin{equation}\label{Omega_gg_omega}
\hbar\Omega\gg \hbar\omega.
\end{equation}
Assuming~(\ref{Omega_gg_omega}) and taking into account (\ref{dipole:chi12}), we obtain for the
dipole moment ${\hatd}$:
\begin{eqnarray}
\nonumber && {\hatd} = \frac{2\pi i  e \hbar}{ m^2 \Omega^2 }
\frac{1}{T}\int_0^T dt e^{in\omega t} \\
&& \times \Big[ L^{(1)}(t)\vk{P}_i(t)
-L^{(2)}(t)\vk{P}_f(t)\Big],\label{hatd}
\end{eqnarray}
where $\vk{P}_{i,f}(t)=\vk{p}_{i,f}-(e/c)\vk{A}(t)$.

The result for the zero-order approximation $\vk{d}^{\dir}_n$ follows from (\ref{hatd}) in three steps: (1) we substitute the zero-order approximation $f_\vp^{\dir}(t)$ [given by
(\ref{f(t):dr})] for $f_\vp(t)$ into the results (\ref{L1}) and~(\ref{L2}) for $L^{(1,2)}(t)$; (2) we
substitute the results obtained for $L^{(1,2)}(t)$ into (\ref{hatd}); (3) the integral over
time in (\ref{hatd}) is evaluated using the saddle point method. As a result, for $\vk{d}^{\dir}_n$
we obtain~(cf.~\cite{ResBrs}):
\begin{equation}
\label{DIPZR}
\vk{d}_{n}^{\dir}(\vk{p}_i,\vk{p}_{f})=i^nJ_{n}(\rho)\,
\vk{d}^{(0)}\left[\vk{P}_i(t_0),\vk{P}_{f}(t_0)\right],
\end{equation}
where $\vk{d}^{(0)}$ is the field-free BrS dipole moment in the effective range approximation:
\begin{eqnarray}
\label{TDER:d0:BrS} && \vk{d}^{(0)}(\vk{p},\vk{p}')=\frac{8 \pi i 
e\hbar^3}{(\vk{p}^2-\vk{p}'^2)^2 }\big[\vk{p}
\mathcal{A}(p')-\vk{p}'\mathcal{A}(p)\big],
\end{eqnarray}
$\rho=eF (\vk{p}_{i}-\vk{p}_{f})\cdot \vk{e}_z/(m\hbar\omega^2)$,
$J_n$ is a Bessel function and the saddle point $t_0$ satisfies
the equation $\rho\sin\omega t_0 = n$ or
$\vk{P}_i^2(t_0)-\vk{P}_f^2(t_0)=2 m \hbar\Omega$.

\subsection{Saddle-point result for the ``rescattering'' part, $\vk{d}^{\resc}_n$, of the LABrS amplitude  }

In order to find the rescattering correction, ${\hatd}^{\resc}$, to the result~(\ref{DIPZR})
originating from the dipole moment ${\hatd}$, we replace the functions $f_{\vk{p}_{i,f}}$ in
expressions~(\ref{L1}) and~(\ref{L2}) for $L^{(1,2)}(t)$ by their rescattering approximations
$f_{\vk{p}_{i,f}}\approx f_{\vk{p}_{i,f}}^{\resc}$ given by~(\ref{f(t):rsc}). The result for
the dipole moment ${\hatd}^{\resc}$ then follows from~(\ref{hatd}) after substituting there the
``rescattering approximations'' for $L^{(1,2)}(t)$. The result can be presented in the form:
\begin{equation}
{\hatd}^{\resc}=\frac{1}{T}\int_0^T dt\Big( \sum_k
\hat{\boldsymbol{\mcal{D}}} + \sum_k
 \hat{\boldsymbol{\mcal{D}}}\big|_{\mcal{G}}\Big),
\label{hatd2}
\end{equation}
where
\begin{eqnarray}\nonumber
 &&\hat{\boldsymbol{\mcal{D}}} =-\frac{2\pi i e \hbar}{
m^2 \Omega^2 }\vk{P}_f(t) \mcal{A}\left[Q(t,t'_k)\right]\\
&& \times \chi_{\vk{p}_i}(t,t'_k) e^{i n \omega
t-iS_{\vk{p}_f}({t})/\hbar},\label{cD}
\end{eqnarray}
and the symbol $\mcal{G}$ in~(\ref{hatd2}) implies the following set of replacements in $\hat{\boldsymbol{\mcal{D}}}$:
\begin{equation}
\label{replacement} \mcal{G}=\{\vk{p}_i\leftrightarrow -\vk{p}_f,
t\to -t, t'_k\to -t'_k, n\to -n\}.
\end{equation}
The first summation (over
$\hat{\boldsymbol{\mcal{D}}}$) in the integrand of~(\ref{hatd2})
is taken over saddle points $t'_k=t'_k(t)$ that satisfy the
equation:
\begin{equation}
\label{ec1} \vk{P}_i^2(t'_k)=\vk{Q}^2(t'_k,t),\quad t'_k<t.
\end{equation}
The second summation in
the integrand of~(\ref{hatd2}) is over $\hat{\boldsymbol{\mcal{D}}}\big|_{\mcal{G}}$\,, which contains
$\chi_{-\vk{p}_f}(-t,-t'_k)$ [cf.~(\ref{cD}) and~(\ref{replacement})]. It involves the saddle points
$t'_k$ satisfying the equation:
\begin{equation}
\label{ec2} \vk{Q}^2(t'_k,t)=\vk{P}_f^2(t'_k),\quad t'_k>t.
\end{equation}
Both equations~(\ref{ec1}) and~(\ref{ec2}) follow from the saddle point equation~(\ref{sp1}).

Let us consider now the rescattering approximation for the dipole
moment $\tld$ given by~(\ref{Scatt-Scatt}). As shown in the
Appendix, $\tld$  can be expressed as the sum of two
terms:
\begin{eqnarray}
&& {\tld}={\tld}^{(+)}+{\tld}^{(-)}, \label{tld:sum} \\
&& {\tld}^{(+)}=\frac{e}{ \Omega^2}\sqrt{\frac{2 \pi i \hbar}{
m^3}}\frac{1}{T}\int_0^Tdt \int\limits_{-\infty}^{t} \frac{dt'
\boldsymbol{\mcal{Q}}(t,t')}{(t-t')^{3/2}}\nonumber
\\
&& \times e^{i n\omega t+(i/\hbar) \left[\epsilon_i(t-t')+
S(t,t')\right]}
f_{\vk{p}_{i}}\left(t'\right)f^{}_{-\vk{p}_{f}}\left(-t\right),
\label{tld+}\\
&& {\tld}^{(-)}=-\frac{e}{\Omega^2}\sqrt{\frac{2 \pi \hbar}{ i
m^3}}\frac{1}{T}\int_0^Tdt \int\limits^{\infty}_{t} \frac{dt'
\boldsymbol{\mcal{Q}}(t,t')}{(t-t')^{3/2}}\nonumber
\\
&& \times e^{i n\omega t+(i/\hbar)\left[\epsilon_f (t'-t)+
S(t',t)\right]}
 f_{\vk{p}_{i}}\left(t\right) f_{-\vk{p}_{f}}\left(-t'\right),
\label{tld-}
\end{eqnarray}
where
\begin{eqnarray}
&&\boldsymbol{\mcal{Q}}(t,t')=\vk{q}(t,t')+ \vk{e}_z
\frac{i e F}{2 \omega^2}\big[ e^{i\omega t} \zeta_{+}
+e^{-i\omega t} \zeta_{-}\big], \label{cQ}  \\
 &&\zeta_{\pm}=\Omega^2/(\Omega\pm \omega)-\Omega.
\label{Omega_xx}
\end{eqnarray}

Note that although each integral ${\tld}^{(\pm)}\equiv \tl{\vk{d}}^{(\pm)}_n(\vk{p}_i,\vk{p}_f)$ is
divergent at $t'=t$, their sum is convergent. Note also the symmetry relation:
$\tl{\vk{d}}^{(-)}_n(\vk{p}_i,\vk{p}_f)=\tl{\vk{d}}^{(+)}_{-n}(-\vk{p}_f,-\vk{p}_i)$, which follows
from~(\ref{tld+}) and~(\ref{tld-}) after changing
the variables of integration
 ($t\rightarrow -t$, $t'\rightarrow -t'$) in
 $\tl{\vk{d}}^{(+)}_{-n}(-\vk{p}_f,-\vk{p}_i)$
 [taking into account the relation $S(t',t)=S(-t,-t')$ for the classical action, and also that in (\ref{Omega_xx}) $\Omega\equiv
\Omega(p_i,p_f,n)=(p_i^2-p_f^2)/(2 m\hbar)+n\omega$, so that $\Omega(p_f,p_i,-n)=-\Omega$]. Hence
the following symmetry relation for the exact LABrS dipole moment is valid:
$\vk{d}_n=\vk{d}_n^{(+)}(\vk{p}_i,\vk{p}_f)+\vk{d}^{(-)}_n(\vk{p}_i,\vk{p}_f)$, where
$\vk{d}^{(-)}_n(\vk{p}_i,\vk{p}_f)=\vk{d}^{(+)}_{-n}(-\vk{p}_f,-\vk{p}_i)$. This relation is in
agreement with the invariance of the matrix element $\vk{d}_n$ with respect to time inversion.

For high-energy LABrS, under the assumption~(\ref{Omega_gg_omega}), we have $\zeta_{\pm}\approx
\mp\omega$ and $\boldsymbol{\mcal{Q}}(t,t') \approx \vk{Q}(t,t')$ [cf.~(\ref{Q}) and~(\ref{cQ})].
Replacing $\boldsymbol{\mcal{Q}}$ by $\vk{Q}$ in (\ref{tld+}) and~(\ref{tld-}),  substituting there
the ``direct'' scattering approximation result~(\ref{f(t):dr}) for $f_{\vp}(t)$ ($f_{\vp}\approx
f_{\vp}^{\dir}$) and using the regular saddle point method for carrying out the integration over $t'$,
we obtain:
\begin{equation}
{\tld}^{\resc}=\frac{1}{T}\int_0^T dt \Big( \sum_k
\tl{\boldsymbol{\mcal{D}}} + \sum_k
\tl{\boldsymbol{\mcal{D}}}\big|_{\mcal{G}}\Big), \label{tld2}
\end{equation}
where
\begin{eqnarray}\nonumber
&&\tilde{\boldsymbol{\mcal{D}}} =\frac{2\pi i e \hbar}{ m^2
\Omega^2 }\vk{Q}(t,t'_k) \mcal{A}(P_f(t))\\
&& \times \chi_{\vk{p}_i}(t,t'_k) e^{i n \omega
t-iS_{\vk{p}_f}({t})/\hbar},\label{cD2}
\end{eqnarray}
and the operation $\mcal{G}$ in the second sum in (\ref{tld2}) is
defined by~(\ref{replacement}). The saddle points $t'_k=t'_k(t)$
in $\tilde{\boldsymbol{\mcal{D}}}$ and
$\tl{\boldsymbol{\mcal{D}}}\big|_{\mcal{G}}$ satisfy respectively equations
(\ref{ec1}) and (\ref{ec2}).

Combining the results~(\ref{hatd2}) and~(\ref{tld2}) [cf. Eq.~(\ref{dip111})], we obtain the integral expression
for the rescattering correction to the ``direct'' LABrS dipole moment~(\ref{DIPZR}):
\begin{equation}
\label{d:resc} \vk{d}^{\resc}_n=\frac{1}{T}\int_0^T dt \Big(\sum_k
\boldsymbol{\mcal{D}}+\sum_k
\boldsymbol{\mcal{D}}\big|_{\mcal{G}}\Big),
\end{equation}
where
\begin{eqnarray}
\label{cD:sum} &&\boldsymbol{\mcal{D}}=
\hat{\boldsymbol{\mcal{D}}}+\tl{\boldsymbol{\mcal{D}}}
=\boldsymbol{\mcal{F}}(t,t'_k(t))e^{i\phi(t,t'_k(t))},\\
\label{phi}
 && \phi(t,t'_k)=n \omega
t-S_{\vk{p}_f}({t})/\hbar+\varphi_{\vk{p}_i}(t,t'_k),\\
\label{pre-faktor} &&\boldsymbol{\mcal{F}}(t,t'_k)=\frac{2\pi i
e\hbar}{ m^2 \Omega^2}\frac{\mcal{A}\left[P_{i}(t'_k)\right]
}{\sqrt{(\hbar/m)(t-t'_k)^3
 \mcal{D}_{\vk{p}_i}(t,t'_k)}}\nonumber\\
 &&\times
 \left[\vk{Q}(t,t'_k)\mcal{A}(P_f(t))-\vk{P}_f(t)\mcal{A}(Q(t,t'_k))\right],
\end{eqnarray}
and where $\mcal{D}_{\vk{p}}(t,t')$ is defined by (\ref{D_def}). The exponential factor, $\exp(i\phi)$, in
(\ref{cD:sum}) is a highly oscillatory function of the time $t$, while the pre-exponential factor
$\boldsymbol{\mcal{F}}$
is a slowly-varying function of $t$. Thus to estimate the integral
over $t$ in (\ref{d:resc}), we use the regular saddle point
method, which gives the following result:
\begin{equation}
\label{d:resc2}
\vk{d}^{\resc}_n=\vk{d}^{\rescI}_n+\vk{d}^{\rescII}_n,
\end{equation}
where $\vk{d}^{\rescI}_n$ and $\vk{d}^{\rescII}_n$ originate from the first and second terms of the
integrand in (\ref{d:resc}) respectively. They are expressed through the field-free quantities
[i.e., the BrS dipole moment $\vk{d}^{(0)}$ given by (\ref{TDER:d0:BrS}) and the elastic electron
scattering amplitude $\mcal{A}$ given by (\ref{scat.ampl})]:
\begin{eqnarray}\label{tdp1}
&&\vk{d}^{\rescI}_n=  \sum_k
\vk{d}^{(0)}\left[\vk{Q}(t_{k},t'_k),\vk{P}_f(t_{k})\right]
\mcal{A}[P_i(t'_{k})] \, \mcal{W}_{k},\\
&& \vk{d}^{\rescII}_n= \sum_k
\vk{d}^{(0)}\left[\vk{P}_i(t_{k}),\vk{Q}(t_{k},t'_k)\right]
\mcal{A}[P_f(t'_{k})] \, \mcal{W}'_{k}.\label{tdp2}
\end{eqnarray}
Here $\mcal{W}_{k}\equiv  \mcal{W}_{\vk{p}_i,\vk{p}_f}(t_k,t'_k)$,
$\mcal{W}'_{k}\equiv  \mcal{W}_{-\vk{p}_f,-\vk{p}_i}(-t_k,-t'_k)$,
\begin{eqnarray}
\label{propag}
&&\mcal{W}_{\vk{p}_i,\vk{p}_f}(t,t') = \sqrt{\frac{ m i }{ 2 \pi
\hbar \mcal{K}}}\frac{ \omega
 e^{i\phi(t,t')}}{\sqrt{
 (t-t')^3 \mcal{D}_{\vk{p}_i}(t,t')}},\\
\nonumber &&\mcal{K}\equiv\mcal{K}_{\vk{p}_i,\vk{p}_f}(t,t') \equiv d^2
\phi(t,t'_{k}(t))/dt^2|_{t'_k(t) = t'}\\ && =-\frac{\left(\vk{Q}(t,t')\cdot
\vk{Q}(t',t)\right)^2}{m^2\hbar^2(t-t')^2 \mcal{D}_{\vk{p}_i}(t,t')} -\mcal{D}_{\vk{p}_f}(t',t).
\label{K(t,t')}
\end{eqnarray}

The times $t_k$ and $t_k'$ in~(\ref{tdp1}) are the $k$th solution of the system of coupled saddle
point equations, (\ref{ec1}) and $[d\phi(t,t'_k(t))/dt]|_{t=t_k}=0$. The latter equation may be
written in the explicit form:
\begin{eqnarray}
&& \label{ecx1} \hbar\Omega=\mcal{E}_1(t_k),
\end{eqnarray}
where
\begin{eqnarray}
&&\label{E1(t)}
\mathcal{E}_1(t)=\frac{\vk{Q}^2(t,t'_k(t))-\vk{P}_f^2(t)}{2m}.
\end{eqnarray}
Similarly, the times $t_k$ and $t_k'$ in
(\ref{tdp2}) are the $k$th solution of the system of equations comprised of equation~(\ref{ec2}) and
the following equation:
\begin{eqnarray}\label{ecx2}
&&\hbar\Omega=\mcal{E}_2(t_k),
\end{eqnarray}
where
\begin{eqnarray}
&&\label{E2(t)}
\mathcal{E}_2(t)=\frac{\vk{P}_i^2(t)-\vk{Q}^2(t,t'_k(t))}{2m}.
\end{eqnarray}
The summation over $k$ in~(\ref{tdp1}) and~(\ref{tdp2}) is taken over pairs of saddle points
$\mathcal{P}_k \equiv (t_k,t_k')$ for which $\mathrm{Im}\,\phi(t_k,t_k')\geq0$.

\subsection{Three-step rescattering interpretation of the results~(\ref{tdp1}) and~(\ref{tdp2})}

The results (\ref{d:resc2}) -- (\ref{tdp2}) for the rescattering part $\vk{d}^{\resc}_n$ of the LABrS
amplitude allow a clear physical interpretation in terms of the rescattering scenario for the LABrS
process. The principal difference of this scenario from that for other collision processes, such as
LAES and LARR/LARA, is that the rescattering LABrS amplitude~(\ref{d:resc2}) involves a sum of two
terms. Although both of these terms have a similar structure (i.e. each is expressed as a sum of factorized
three-term products involving two field-free quantities, $\vk{d}^{(0)}$ and $\mcal{A}$), the terms
$\vk{d}^{\rescI}_n$ and $\vk{d}^{\rescII}_n$ have different rescattering interpretations. Thus the
entire rescattering picture for the LABrS process is more complicated than are those for LAES or LARR/LARA
since two different rescattering scenarios (scenario I and scenario II) for LABrS can be formulated
in accordance with the results~(\ref{tdp1}) and~(\ref{tdp2}).

\subsubsection{The rescattering scenario I.}
The $k$-th term of the sum in (\ref{tdp1}) describes the following
picture (the rescattering scenario~I). The electron with a given
initial momentum $\vp_i$ elastically scatters from the potential
$U(r)$ at the time moment $t_k'$. This first step of the
rescattering scenario is described by the amplitude
$\mathcal{A}[P_i(t'_k)]$ for elastic field-free scattering. Since
the collision takes place in the presence of a field ${\bf F}(t)$,
the amplitude $\mathcal{A}$ involves the laser-modified
instantaneous momentum $\vk{P}_i(t_k')$ of the electron at the
moment of collision (instead of the momentum $\vp_i$). The
scattering direction is given by the vector $\vk{Q}(t_k',t_k)$,
which is determined only by the vector potential of the laser
field and has the sense of an intermediate ``kinetic momentum'' of
the electron in an ``intermediate'' state, immediately after the
elastic scattering event at the moment $t_k'$. The amplitude
$\mathcal{A}[P_i(t'_k)]$ (involving the instantaneous momentum
$\vP_i$) describes the elastic scattering [since $|\vQ|=|\vP_i|$
in accordance with equation~(\ref{ec1})], while the initial
momentum $\vp_i$ changes to $\vq$ ($|\vp_i|\neq |\vq|$). From this
intermediate state, the electron starts to move in the laser field
up to the moment of the second scattering (or rescattering). One
must thus ensure that the electron returns back to the origin
[where the magnitude of the potential $U(r)$ is largest] at the
moment $t_k$. Now the momentum vector $\vq = \vq(t_k,t_k')$ [cf.
the definition~(\ref{q})] depends on both times, the time of the
first collision ($t'_k$) and the time of the rescattering ($t_k$).
This pair of times, $\mathcal{P}_k = (t_k,t_k')$, corresponds to a
closed classical trajectory of the free electron's motion in the
laser field $\vF(t)$.  The momentum $\vq(t_k,t_k')$ can be found
by solving the classical equations for an electron in the field
$\vF(t)$ with the boundary conditions $\vr(t_k)=\vr(t'_k)=0$ (cf.,
e.g.,~\cite{jetpl02}). The electron's motion along the $k$th
trajectory during the time interval $\Delta t_k=t_k-t_k'$ is the
second step of the rescattering scenario and this step is
described by the ``propagation'' factor $\mcal{W}_k$ in
(\ref{tdp1}). During this motion the electron gains (or loses)
energy from the laser field and changes its intermediate kinetic
momentum from $\vk{Q}(t'_k,t_k)$ to $\vk{Q}(t_k,t'_k)$. As a
result of the rescattering at the moment $t_k$, the electron with
the initial (i.e., intermediate) momentum $\vq$ rescatters along
the direction of the final asymptotic momentum $\vp_f$. The
rescattering event (the third step) is accompanied by emission of
a spontaneous photon $\hbar\Omega$. The electron thus changes its
kinetic momentum $\vk{Q}(t_k,t_k')$ to the instantaneous momentum
$\vk{P}_f(t_k)$, so that the kinetic energy decreases by the value
$\mathcal{E}_1(t_k) = \hbar\Omega$ in accordance with
equations~(\ref{ecx1}) and~(\ref{E1(t)}). The third step of the
rescattering scenario~I is described by the field-free BrS dipole
moment $\vk{d}^{(0)}[\vk{Q}(t_k,t_k'),\vk{P}_f(t_k)]$. The
summation over $k$ in equation~(\ref{tdp1}) results in
interference of the partial rescattering LABrS amplitudes
corresponding to the different classical trajectories that are
related to the saddle points $\mathcal{P}_k$. As may be seen from
the expression~(\ref{propag}), the propagation factor
$\mathcal{W}_k$ is proportional to the spreading factor $\Delta
t_k^{-3/2}$, so that as the corresponding electron travel time
increases, the contribution of the $k$th term decreases.

\subsubsection{The rescattering scenario II.}
The result~(\ref{tdp2}) describes the second rescattering scenario (scenario~II). The difference
from scenario~I is that the spontaneous photon emission occurs at time $t_k$
of the first scattering (the first step). At this time the electron's kinetic energy decreases
by the value $\mathcal{E}_2(t_k) = \hbar\Omega$ in accordance with the saddle point
equation~(\ref{ecx2}). The second step is the electron's motion along the $k$th closed classical trajectory
during the period $\Delta t_k=t_k'-t_k$; this is described by the factor $\mathcal{W}'_k$.
The third step is the elastic electron-atom scattering with instantaneous momentum
$\vk{P}_f(t_k')$ at the time $t_k'$. The scenario~II may occur for an arbitrarily high energy $E_i$
of the incident electron: at the first electron-atom collision $E_i$ decreases by spontaneous
emission, so that the laser field may return the slow electron back to the atom. This fact is in
contrast with the rescattering scenario~I with elastic scattering at the first step, for which the
laser field returns electrons back to the atom only for $E_i< 10 u_p$. (This latter condition is discussed
in~\cite{jetpl02} for the LAES process, which allows the three-step rescattering interpretation
with the same first step of elastic scattering.) Note that for $\Omega=0$ each of the two systems
of saddle point equations~(\ref{ec1}),~(\ref{ecx1}) or~(\ref{ec2}),~(\ref{ecx2}) (for the
scenarios~I and~II
respectively) reduces to the same saddle point equations describing the
rescattering scenario for the case of LAES process~\cite{LAES2013}. The system
(\ref{ec1}),~(\ref{ecx1}) reduces to
\begin{eqnarray} \label{saddleLAES}
&&\vk{P}_i^2(t'_k)=\vk{Q}^2(t'_k,t_k),\nonumber \\
&& \vk{Q}^2(t_k,t'_k)=\vk{P}_f^2(t_k),
\end{eqnarray}
while the result for ~(\ref{ec2}),~(\ref{ecx2}) coincides with (\ref{saddleLAES}) upon making the substitutions
$t_k\leftrightarrow t'_k$.

\subsubsection{Extent of the rescattering plateau in the $\Omega$ dependence of the LABrS cross section.}

The key factors in the dipole moments~(\ref{tdp1}) and~(\ref{tdp2}) are the propagation factors
$\mcal{W}_{k}$ and $\mcal{W}_{k}'$, which describe the plateau-like behavior of the LABrS spectra. They
depend on the emitted photon energy $\hbar\Omega$ through the $\Omega$ dependence of the saddle
points $\mathcal{P}_k$. The averaged slow $\Omega$ dependence of the LABrS amplitude and cross
section occurs for real solutions  $\mathcal{P}_k$ of equations~(\ref{ec1}) and~(\ref{ecx1}), for
which $ \mcal{W}_{k}$ oscillates [or for real solutions of equations~(\ref{ec2}) and~(\ref{ecx2}),
for which $\mcal{W}_{k}'$ oscillates]. For $\hbar\Omega$ larger than the global maximum
$\hbar\Omega_{\max}$ of both functions $\mathcal{E}_{1}(t)$ and $\mathcal{E}_{2}(t)$, all saddle
points $\mathcal{P}_k$ become complex and the LABrS amplitude decays exponentially, so that $\hbar\Omega_{\max}$ 
determines the position of the rescattering plateau cutoff.

\begin{figure}
\begin{center}
\includegraphics[width=1.\linewidth]{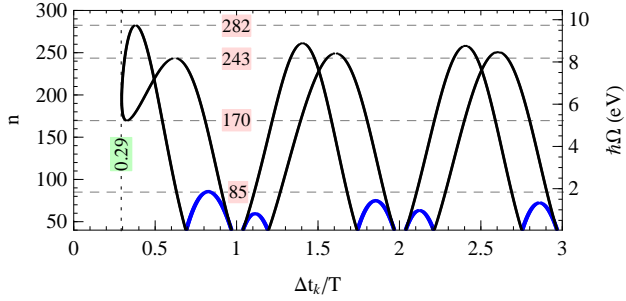}
\end{center}
\caption{
The dependence of $\hbar\Omega$ ($ = E_i-E_f+n\hbar\omega$) on the electron's travel time $\Delta t_k=|t_k-t_k'|$
along various closed trajectories for a laser field with $I=1.06\times 10^{10}$ W/cm$^2$, $\hbar\omega=0.04$ eV, $T=2\pi/\omega$ and for initial and final electrons having $\vp_i\parallel \vp_f\parallel \ve_z$, $E_i=3.47$ eV, $E_f=5.03$ eV. The thick blue lines are for the rescattering scenario I, $\hbar\Omega=\mathcal{E}_{1}(t_k)$; the thin black lines are for the rescattering scenario II, $\hbar\Omega=\mathcal{E}_{2}(t_k)$. The vertical line indicates the minimal travel time,
corresponding to the shortest closed trajectory; the horizontal lines indicate some local extrema of
$\mathcal{E}_{1,2}(t_k)$ with $\Delta t_k<T$.}
\label{Omega:t}
\end{figure}

Figure~\ref{Omega:t} shows the BrS photon energy $\hbar\Omega=\mathcal{E}_{1,2}(t_k)$ as a function
of the electron's travel time $\Delta t_k=|t_k-t_k'|$ along various closed trajectories in accordance
with the previously-described rescattering scenarios. A two-valued dependence of $\mathcal{E}_{2}(t_k)$
on $\Delta t_k$ is due to the character of solutions of the saddle point equation~(\ref{ec2}),
i.e.~the numerical analysis of this equation shows that we have two rescattering times $t'_k(t)$ for a given time $t$ of the first scattering event.
As may be seen in figure~\ref{Omega:t}, while $\mathcal{E}_{1}(t_k)$ is a single-valued function of
$\Delta t_k$, the dependence of $\mathcal{E}_{2}(t_k)$ on $\Delta t_k$ exhibits the single local
minimum ($n=170$) near the shortest closed trajectory with $\Delta t_k\approx 0.29T$
($T=2\pi/\omega$) and a set of local maxima. The first maximum ($n=282$ or
$\hbar\Omega_{\max}=9.74$~eV) at $\Delta t_k\approx 0.38T$ is a global maximum of
$\mathcal{E}_2(t_k)$ and it determines the upper bound of $\Omega$, at which real solutions
$\mathcal{P}_k$ of the equations~(\ref{ec2}) and~(\ref{ecx2}) (corresponding to closed classical
trajectories) exist. With decreasing $n$, the number of closed classical trajectories increases by
the addition of two real solutions each time one crosses a local maximum of
$\mathcal{E}_2(t)$. We expect that the trajectories related to the first (global) and second (at
$\Delta t_k\approx 0.61T$) maxima should contribute to the LABrS amplitude significantly, while all
other trajectories corresponding to local maxima for $\Delta t_k > T$ contribute less.
The scenario I (with spontaneous photon emission 
during the rescattering event) describes the plateau structure
for $\hbar\Omega$ smaller than the global maximum of
$\mathcal{E}_1(t_k)$ at $\Delta t_k=0.84T$, i.e. only for
$n\leqslant 85$ or $\hbar\Omega<1.86$~eV.

\begin{figure*}
\begin{center}
\includegraphics[width=\linewidth]{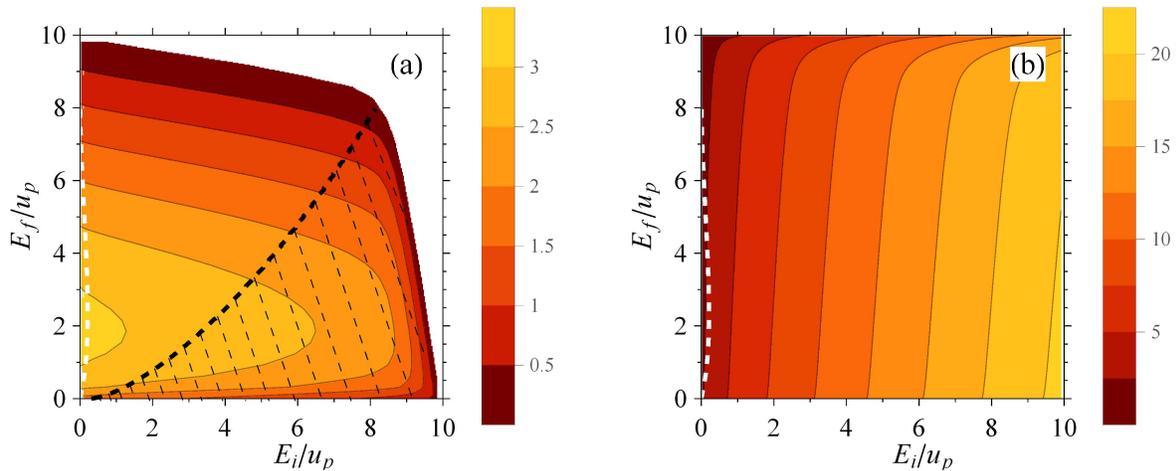}
\end{center}
\caption{ (Color online) Maxima of the emitted energy (in units of
$u_p$) in accordance with the rescattering scenarios I
[$\max\mathcal{E}_1$ in panel (a)] and II [$\max\mathcal{E}_2$ in
panel (b)] as a function of initial, $E_i$, and final, $E_f$,
electron energies for the parallel geometry,
$\vp_i\|\vp_f\|\ve_z$. The black dashed line indicates the
boundary of the region (indicated by the shaded area below this
line) for which the cutoff position of the ``direct'' LABrS
plateau is greater than $\max\mathcal{E}_1$. The white dashed
lines in (a) and (b) indicate the boundary of the region (left of
the line) in which $\max\mathcal{E}_1>\max\mathcal{E}_2$. }
\label{E12}
\end{figure*}

In order to estimate the relative contributions of the
rescattering amplitudes $\vd_n^{\rescI}$ and $\vd_n^{\rescII}$ for
the rescattering scenarios I and II for different initial and
final electron energies, in figure~\ref{E12} we compare the global
maxima $\max\mathcal{E}_{1,2}$ of the emitted energies
$\mathcal{E}_{1}(t)$ and $\mathcal{E}_{2}(t)$ corresponding to
scenarios~I and~II respectively. As figure~\ref{E12} shows, the
scenario~I [panel (a)] considerably contributes over a bounded
region of initial ($E_i$) and final ($E_f$) electron energies
limited by the value $10u_p$. Moreover, in the region $E_f\lesssim
E_i$ the amplitude $\vd_n^{\dir}$ for the direct LABrS exceeds the
amplitude $\vd_n^{\rescI}$ for the rescattering term, i.e., for
this region the maximum classically-allowed energy
$\hbar\Omega^{\dir}_{\max}$ for the direct process~\cite{ResBrs}
exceeds the value $\max\mathcal{E}_1$. The maximal
classically-allowed emitted energy $\max\mathcal{E}_1\approx
3.16u_p$ in accordance with the scenario I appears for low
energies $E_i\to 0$ and for $E_f\approx 1.82u_p$. For the scenario
II in figure~\ref {E12}(b), the initial energy $E_i$ is unlimited,
while $E_f<10u_p$. The maximal emitted energy $\max\mathcal{E}_2$
increases with increasing $E_i$ and achieves greater values than
for the case of scenario~I. Only in the tiny region $E_i\lesssim
0.2\up$, $E_f\lesssim 8\up$ (bounded by the dashed white lines in
figure~\ref{E12}) does the amplitude $\vd_n^{\rescI}$ exceed that
of $\vd_n^{\rescII}$.

\subsection{Contributions of coalescing trajectories to the LABrS amplitude}

Local extrema of the functions $\mathcal{E}_{1,2}(t)$ [defined in
equations~(\ref{E1(t)}) and~(\ref{E2(t)})] at $t=\ov{t}_\nu$
($\nu=1,2,\ldots$) correspond to critical values of the emitted
photon energy $\hbar\ov{\Omega}_{\nu}$. For example, let
$\max[\mathcal{E}_j(t)]_\nu=\hbar\ov\Omega_{\nu}$ ($j=1$ or~2) be
the $\nu$th local maximum of the function $\mathcal{E}_j(t)$
(similar reasoning may be used for a local minimum), and let
$\mathcal{P}_{k_1}$ and $\mathcal{P}_{k_2}$ be the two nearest
real saddle points for $\Omega\lesssim \ov{\Omega}_{\nu}$, cf.
figure~\ref{Omega:t}. These saddle points correspond to two closed
electron trajectories in the laser field, which coalesce at
$\Omega\to\ov\Omega_{\nu}$: $\mathcal{P}_{k_1}\to
\ov{\mathcal{P}}_\nu$, $\mathcal{P}_{k_2}\to \ov{\mathcal{P}}_\nu$
[where $\ov{\mathcal{P}}_\nu = (\ov{t}_\nu,\ov{t'}_\nu)$ is the
point of coalescence]. For $\Omega>\ov\Omega_{\nu}$, the
considered saddle points $\mathcal{P}_{k_1}$ and
$\mathcal{P}_{k_2}$ become complex. The merging of two
trajectories $\mathcal{P}_{k_1}$ and $\mathcal{P}_{k_2}$ at
extrema of the emitted energies $\mcal{E}_{1,2}(t)$ is related to
the vanishing of the second derivative~(\ref{K(t,t')}) of the
phase function $\phi(t,t_k'(t))$:
\begin{eqnarray}
\label{K=0:1}
 &&\mcal{K}_{\vk{p}_i,\vk{p}_f}(t,t'_k(t)) =0,\\
\label{K=0:2}
 &&\mcal{K}_{-\vk{p}_f,-\vk{p}_i}(-t,-t'_k(t)) =0.
\end{eqnarray}
Equations~(\ref{K=0:1}) and (\ref{K=0:2}) correspond to the
results~(\ref{tdp1}) for $\vk{d}_n^{\rescI}$ and~(\ref{tdp2}) for
$\vk{d}_n^{\rescII}$ respectively. Hence, the merging point
$\ov{\mathcal{P}}_\nu$ is a solution of the coupled system of
equations~(\ref{ec1}) and~(\ref{K=0:1}) [or~(\ref{ec2})
and~(\ref{K=0:2})].

\begin{figure*}
\begin{center}
\includegraphics[width=1.\linewidth]{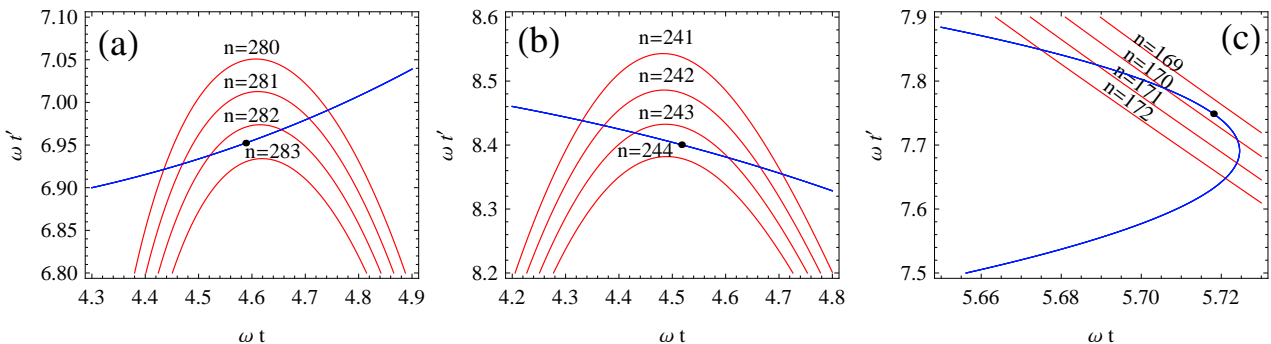}
\end{center}
\caption{(Color online) The saddle point trajectories in the plane
of times ($t,t'$) for the rescattering scenario II. The crossings
of the blue and red lines give the pairs of saddle points
$\mathcal{P}_k=(t_k,t'_k)$. The parameters of the initial and
final electrons and of the laser field are the same as in
figure~\ref{Omega:t}. Trajectories for $\hbar\Omega$ near two
maxima of $\mathcal{E}_2(t_k)$: (a) $n=282$ ($\hbar\Omega=
9.74$~eV) and (b) $n=243$ ($\hbar\Omega= 8.2$~eV); (c)
trajectories near the local minimum of $\mathcal{E}_2(t_k)$ at $n=
170$ ($\hbar{\Omega}= 5.22$~eV). The blue and red lines indicate
solutions of the saddle point equations~(\ref{ec2})
and~(\ref{ecx2}) respectively. The black points indicate the
merging points $\ov{\mathcal{P}}_\nu$ [which are solutions of
equations~(\ref{ec2}) and~(\ref{K=0:2})] for two coalescing saddle
points.} \label{traj}
\end{figure*}

In figure~\ref{traj} we illustrate the motion of the saddle points $\mathcal{P}_k$ on the plane
$(t,t')$ with varying $\hbar\Omega$ (or, equivalently, $n$) for the fixed initial $E_i$ and final
$E_f$ electron energies for the case of rescattering scenario II. The blue and red lines in
figure~\ref{traj} indicate solutions of equations~(\ref{ec2}) and~(\ref{ecx2}) respectively, while
their crossing points are the saddle points $\mathcal{P}_k$. As shown in figures~\ref{traj}(a)
and \ref{traj}(b), these points move toward each other with increasing $n$, which corresponds to
the approach of the function $\mathcal{E}_2(t)$ to its local maxima in the vicinity of $n=282$
[figure~\ref{traj}(a)] and $n=243$ [figure~\ref{traj}(b)]. Figure~\ref{traj}(c) illustrates the merging of the pair of
saddle points with decreasing $n$ down to $n=170$ and the approach of the local minimum of
$\mathcal{E}_2(t)$ at $\omega t\approx 5.72$. In the vicinity of this time, the solution $t'_k(t)$
of (\ref{ec2}) (the blue line) exhibits a branch point (at $t=\tilde{t}$\,), which means that the
second partial derivative of the phase function $\varphi_\vp(t,t')$ [given by
expression~(\ref{phi_def})] over $t'$ vanishes at this point, i.e.,
$\mcal{D}_{\vk{p}}(\tilde{t},t'_k(\tilde{t}\,))=0$ [$\mathcal{D}_\vp(t,t'_k(t)) \sim
(dt'_k/dt)^{-1}$]. For coalescing saddle points, the saddle point analysis used to obtain the
rescattering expansion for the function $f_\vp(t)$ [cf.~equation~(\ref{f(t):sum})] is not
applicable and leads to incorrect results.

To take into account the coalescence of the $k_1$th and $k_2$th saddle points (and hence to apply
our results for the vicinity of the critical value $\overline{\Omega}_\nu$ of the BrS photon
frequency $\Omega$), we modify the regular saddle point method for analysis of the integral
in~(\ref{d:resc}) by making use of the general idea of the uniform approximation~\cite{Bleistein,Wong}.
We need only evaluate the integral of the first term ($\boldsymbol{\mathcal{D}}$) in~(\ref{d:resc}),
since the result for the integral of the second term ($\boldsymbol{\mathcal{D}}|_{\mathcal{G}}$)
can be found by replacing the parameters~(\ref{replacement}) in the result obtained for
$\boldsymbol{\mathcal{D}}$. We approximate the phase function $\phi(t,t'_k(t))$ in (\ref{phi}) in
the vicinity of $t=\ov{t}_\nu$ by the following cubic polynomial:
\begin{equation}
\label{phi:cub} \phi(t,t'_k(t)) = \ov{\phi} + \phi_1
\omega(t-\ov{t}_\nu) + (\phi_2/3) [\omega(t-\ov{t}_\nu)]^3,
\end{equation}
where $\ov{\phi}\equiv \phi(\ov{t}_\nu,\ov{t'}_\nu)$ and the coefficients $\phi_1$ and $\phi_2$ are
expressed in terms of the first and third derivatives of the function $\phi(t,t'_k(t))$ over $t$
evaluated at the point $t=\ov{t}_\nu$ [where we denote $\ov{t'}_\nu \equiv t_k'(\ov{t}_\nu)$]: $\phi_1=d\phi/(\omega
dt)|_{t=\ov{t}_\nu}$, $\phi_2=d^3\phi/(2\omega^3dt^3)|_{t=\ov{t}_\nu}$.  The explicit expressions
for $\phi_1$ and $\phi_2$ are:
\begin{eqnarray}
\label{diff1}
&&\phi_1 = \frac{\Omega-\ov{\Omega}_\nu}{\omega},\quad
\ov{\Omega}_\nu = \frac{\vk{Q}^2(\ov{t}_\nu,\ov{t'}_\nu)-\vk{P}_f^2(\ov{t}_\nu)}{2m\hbar}, \\
\label{diff3} &&\phi_2 = \frac{2\up}{\hbar e F} \vk{e}_z\cdot
[\vk{p}_{f}-\vk{q}(\ov{t}_\nu,\ov{t'}_\nu)]\left(\mcal{B}-
  \sin \omega \ov{t}_\nu\right),\label{eta_1}
  \\
&&  \mcal{B}=\frac{\left[(\vk{e}_z\cdot \vk{Q}(\ov{t'}_\nu,\ov{t}_\nu))^3\sin \omega
  \ov{t'}_\nu+\mcal{C}\right] \cos^2\omega \ov{t}_\nu}{\beta (\vk{e}_z\cdot \vk{Q}(\ov{t}_\nu,\ov{t'}_\nu))^3\cos^2\omega \ov{t'}_\nu},
  \nonumber
  \\\nonumber
&& \mcal{C}= \frac{3 e F}{\omega}\big[\vk{Q}(\ov{t'}_\nu,\ov{t}_\nu)\cdot \vk{P}_i(\ov{t'}_\nu) \\
\nonumber
&&\phantom{C} -\beta \vk{Q}(\ov{t}_\nu,\ov{t'}_\nu)\cdot \vk{P}_f(\ov{t}_\nu)
  \big] \cos^2\omega \ov{t'}_\nu,\nonumber \\ \nonumber
&& \beta= \left(\frac{ \vk{e}_z\cdot [\vk{p}_{i}-\vk{q}(\ov{t}_\nu,\ov{t'}_\nu)]}
{ \vk{e}_z\cdot [\vk{p}_{f}-\vk{q}(\ov{t}_\nu,\ov{t'}_\nu)]}\right)^2 \frac{\cos\omega\ov{t'}_\nu}{\cos\omega \ov{t}_\nu}.
\end{eqnarray}
In the vicinity of the point $\ov{t}_\nu$, we approximate the pre-exponential factor
$\boldsymbol{\mcal{F}}$ in (\ref{cD:sum}) by the linear polynomial:
\begin{equation}
\label{pre-exp:lin} \boldsymbol{\mcal{F}}(t,t'_k(t)) =
\boldsymbol{c}_0+\boldsymbol{c}_1 \omega(t-\ov{t}_\nu).
\end{equation}
The coefficients $\boldsymbol{c}_{0}$ and $\boldsymbol{c}_{1}$ are
expressed in terms of the values of the function
$\boldsymbol{\mcal{F}}(t,t')$ taken at two close saddle points
$t_{k_1}$, $t_{k_2}$
[$\boldsymbol{\mcal{F}}_{k} \equiv \boldsymbol{\mcal{F}}(t_k,t'_k)$,
$k=k_1,k_2$]:
\begin{eqnarray*}
\label{vc0}
  && \boldsymbol{c}_0=\frac{(t_{k_2}-\ov{t}_\nu)\boldsymbol{\mcal{F}}_{k_1}-
 (t_{k_1}-\ov{t}_\nu) \boldsymbol{\mcal{F}}_{k_2}}{t_{k_2}-t_{k_1}}, \\
\label{vc1}
  && \boldsymbol{c}_1=\frac{\boldsymbol{\mcal{F}}_{k_2}-\boldsymbol{\mcal{F}}_{k_1}}{\omega(t_{k_2}-t_{k_1})}.
\end{eqnarray*}
Substituting the expressions~(\ref{phi:cub}) and~(\ref{pre-exp:lin}) into (\ref{cD:sum}) and
extending the range of integration over $t$ to $(-\infty,\infty)$, the integral in (\ref{d:resc})
can be evaluated analytically in terms of the Airy function $\m{Ai}(x)$  and its derivative
$\m{Ai}'(x)$:
\begin{eqnarray}
\nonumber
&&\frac{1}{2\pi}\int_{-\infty}^{\infty}dx
(\boldsymbol{c}_0+\boldsymbol{c}_1 x)
e^{i ( \ov{\phi} + \phi_1 x + \phi_2 x^3/3 ) }\\
&& =e^{i\ov{\phi}}\eta\Big[\boldsymbol{c}_0\m{Ai}\left( \sg \eta \phi_1\right) -i \boldsymbol{c}_1
\sg \eta\m{Ai}'\left(\sg\eta \phi_1\right)\Big],
\label{Ai_int}
\end{eqnarray}
where $\eta=|\phi_2|^{-1/3}$, $\sg=\m{sgn}(\phi_2)$, $\m{sgn}(x)=\pm1$ for $x\gtrless 0$. Finally,
the corrected result for the contribution $\vk{d}^{\rescI}_{n,\nu}$ of two real coalescing saddle
points with $k=k_1$ and~$k_2$ [which are involved in the rescattering LABrS dipole moment $\vk{d}^{\rescI}_n$
in (\ref{tdp1})] can be presented in the form:
\begin{eqnarray}
\nonumber \vk{d}^{\rescI}_{n,\nu} &=& \frac{\eta
e^{i\ov{\phi}}}{\omega(t_{k_2}-t_{k_1})}
\sum_{j=0,1} (-1)^j \boldsymbol{\mcal{F}}_{k_{j+1}} \\
&\times& \left[
\omega(t_{k_{2-j}}-\ov{t}_\nu)\mathrm{Ai}(\xi_\nu)+i\hat{s}\eta\mathrm{Ai}'(\xi_\nu)\right],
\label{resc:two_terms}
\end{eqnarray}
where
\begin{equation}
\label{Airy:arg}
  \xi_\nu=\sg \eta(\Omega-\ov{\Omega}_\nu)/\omega.
\end{equation}

Approximating the merging point $\ov{t}_\nu$ in (\ref{resc:two_terms}) by the center of the
interval $(t_{k_1},t_{k_2})$, $\ov{t}_{\nu}\approx (t_{k_1}+t_{k_2})/2$ (which is a reasonable approximation in the vicinity of local maxima of the functions $\mathcal{E}_{1,2}(t)$,
 cf.~figures~\ref{Omega:t} and~\ref{traj}), for $\vk{d}^{\rescI}_{n,\nu}$ we obtain
\begin{eqnarray}
\vk{d}^{\rescI}_{n,\nu} &=& \sum_{j=1}^{2}
\vk{d}^{(0)}\big[\vk{Q}(t_{k_j},t_{k_j}'),\vk{P}_f(t_{k_j})\big]
\nonumber \\
 &\times& \mcal{A}\big[P_i(t'_{k_j})\big] \mcal{W}_{k_j}^{(\nu)},
\label{resc:approx}
\end{eqnarray}
where the propagation factors $\mcal{W}_{k_j}^{(\nu)}$ have the following form:
\begin{eqnarray}
\nonumber
\mathcal{W}_k^{(\nu)} &=& \frac{\eta e^{i\ov{\phi}}}{ 2\sqrt{(\hbar/m)({t}_{k}-{t'}_{k})^3
\mcal{D}_{\vk{p}_i}({t}_{k},{t'}_{k})} } \\
&\times& \left[\m{Ai}(\xi_\nu)-\frac{i \sg \eta }{\omega(t_k-\ov{t}_\nu)} \m{Ai}'(\xi_\nu)\right].
\label{propag:W:Ai}
\end{eqnarray}
The result for $\vk{d}^{\rescII}_{n,\nu}$ follows from the results~(\ref{resc:approx})
and~(\ref{propag:W:Ai}) by the following set of substitutions: $\mathcal{G}=\{ \vp_i\leftrightarrow
-\vp_f, \mathcal{P}_k\to -\mathcal{P}_k, \ov{\mathcal{P}}_\nu\to -\ov{\mathcal{P}}_\nu \}$
[cf.~(\ref{replacement})], where the saddle points $\mathcal{P}_k = (t_k,t_k')$ satisfy the system
of equations~(\ref{ec2}) and~(\ref{ecx2}) and the merging point
$\ov{\mathcal{P}}_\nu=(\ov{t}_\nu,\ov{t'}_\nu)$ is a solution of the system of
equations~(\ref{ec2}) and~(\ref{K=0:2}):
\begin{eqnarray}
\nonumber
\vk{d}^{\rescII}_{n,\nu} &=&  \sum_{j=1}^{2}
\vk{d}^{(0)}\big[\vk{P}_i(t_{k_j}),\vk{Q}(t_{k_j},t_{k_j}')\big] ,
\\
&\times& \mcal{A}[P_f(t'_{k_j})]\mcal{W\,'}_{k_j}^{(\nu)},
\label{resc2:approx}
\end{eqnarray}
where $\mathcal{W\,'}_k^{(\nu)} = \mathcal{W}_k^{(\nu)}\big|_{\mathcal{G}}$.

The result~(\ref{resc:approx}) [or~(\ref{resc2:approx})] describes the contributions of two real
coalescing saddle points $\mathcal{P}_{k_1}$ and $\mathcal{P}_{k_2}$, corresponding to two terms in
(\ref{tdp1}) [or (\ref{tdp2})] with $k=k_1,k_2$. As discussed above, these two solutions
become complex for $\hbar\Omega$ greater than the local maximum of the function $\mathcal{E}_1(t)$
[or $\mathcal{E}_2(t)$] or for $\hbar\Omega$ smaller than the local minimum of $\mathcal{E}_{1}(t)$
[or $\mathcal{E}_2(t)$], while in~(\ref{tdp1}) [or~(\ref{tdp2})] we take into account only one of
them, which describes the decrease of the partial LABrS amplitude as $\Omega$ varies. In order
to obtain the result for this term in the vicinity of the critical value $\ov{\Omega}_\nu$,
we first simplify the results~(\ref{resc:approx}) and (\ref{resc2:approx}). We first approximate
the pre-exponential factor $\boldsymbol{\mcal{F}}(t,t')$ by its value at $t=\ov{t}_\nu$\,, i.e. we
use equation~(\ref{pre-exp:lin}) with $\boldsymbol{c}_0=\boldsymbol{\mcal{F}}(\ov{t}_{\nu},
\ov{t'}_\nu)$ and $\boldsymbol{c}_1=0$. Using then~(\ref{Ai_int}), we obtain that the two terms in the
sum for $\vk{d}_{n,\nu}^{\rescI}$ in (\ref{resc:approx}) [and similarly for $\vk{d}_{n,\nu}^{\rescII}$ in (\ref{resc2:approx})] can be replaced by only one term with $\mathcal{P}_{k_1}=\mathcal{P}_{k_2}=\ov{\mathcal{P}}_{\nu}$:
\begin{eqnarray}
&& \vk{d}^{\rescI}_{n,\nu} =
\vk{d}^{(0)}\big[\vk{Q}^{(\nu)},\vk{P}_f^{(\nu)}\big]
\mcal{A}\big[P_i'^{(\nu)}\big] \mcal{W}^{(\nu)},
\label{resc1:1term} \\
&& \vk{d}^{\rescII}_{n,\nu} =
\vk{d}^{(0)}\big[\vk{P}_i^{(\nu)}, \vk{Q}^{(\nu)}\big]
\mcal{A}\big[P_f'^{(\nu)}\big] \mcal{W\,'}^{(\nu)},
\label{resc2:1term}
\end{eqnarray}
where $\vk{Q}^{(\nu)}=\vk{Q}(\ov{t}_{\nu},\ov{t'}_{\nu})$,
$\vk{P}_{i,f}^{(\nu)}=\vk{P}_{i,f}(\ov{t}_{\nu})$, $P_{i,f}'^{(\nu)}=P_{i,f}(\ov{t'}_{\nu})$ and the propagation factors are
\begin{eqnarray}
 &&\mathcal{W}^{(\nu)} = \frac{\ov{\Omega}_{\nu}^2
 \eta e^{i\ov{\phi}}\m{Ai}(\xi_\nu)}{\Omega^2\sqrt{(\hbar/m)(\ov{t}_{\nu}-\ov{t'}_{\nu})^3
\mcal{D}_{\vk{p}_i}(\ov{t}_{\nu},\ov{t'}_{\nu})} },\\
&&  \mathcal{W\,'}^{(\nu)}=\mathcal{W}^{(\nu)}|_{\mathcal{G}}.
\end{eqnarray}
The simplified results~(\ref{resc1:1term}) and
(\ref{resc2:1term}) 
depend only on the real merging point $\ov{\mathcal{P}}_\nu$, and thus they can be used in the
classically-forbidden region of $\Omega$ (i.e., where the saddle points $\mathcal{P}_{k}$ become
complex). We emphasize that the atomic factors $\mcal{A}$ and $\vk{d}^{(0)}$ in
(\ref{resc1:1term}) and (\ref{resc2:1term}) depend on the real kinetic momenta taken at the real times
$\ov{t}_\nu$ and $\ov{t'}_{\nu}$.

For an accurate description of the high-energy (or rescattering) part of the LABrS spectrum in the
low-frequency approximation, we use the following rules: (i) For the interval
$x_0<\eta(\Omega-\ov{\Omega}_\nu)/\omega<0$, where $x_0\approx-2.34$ is the first zero of the Airy
function for negative arguments ($\mathrm{Ai}(x_0)=0$), we replace two terms in the
expression~(\ref{tdp1}) [(\ref{tdp2})] with coalescing saddle points at local maxima
$\hbar\ov{\Omega}_\nu$ of the function $\mathcal{E}_1(t)$ [$\mathcal{E}_2(t)$] by the
result~(\ref{resc:approx}) [(\ref{resc2:approx})]; (ii) For $\Omega>\ov{\Omega}_\nu$ (i.e. for energies
$\hbar\Omega$ exceeding a local maximum of $\mathcal{E}_{1,2}$), we replace a term in~(\ref{tdp1})
[(\ref{tdp2})], originating from the coalescence of two real trajectories, by the
result~(\ref{resc1:1term}) [(\ref{resc2:1term})]; (iii) In the vicinity of the local minimum of the
function $\mathcal{E}_2(t)$ (cf. discussions of figures~\ref{Omega:t} and~\ref{traj}) we extend the
use of the simplified result~(\ref{resc2:1term}) into the region of the singularity of the function
$f_{-\vp_f}(t)$ ($\mathcal{D}_{-\vp_f}\approx 0$); (iv) Except for these special cases (i-iii), the
rescattering results~(\ref{tdp1}) and~(\ref{tdp2}) (obtained in the nonresonant low-frequency approximation) are
used. Note that the partial LABrS amplitudes in expressions~(\ref{tdp1}) and~(\ref{tdp2})
corresponding to complex saddle points $\mathcal{P}_k$ (far away from a coalescence point) can be
evaluated directly [without simplification to the forms~(\ref{resc1:1term})
and~(\ref{resc2:1term})] using the analytic functions $\mathcal{A}$ and $\vd^{(0)}$ obtained within
the effective range approximation.

\section{Numerical results and discussion}

\subsection{Comparisons with exact TDER results and discussions of interference phenomena}

\begin{figure*}
\begin{center}
\includegraphics[width=1.\linewidth]{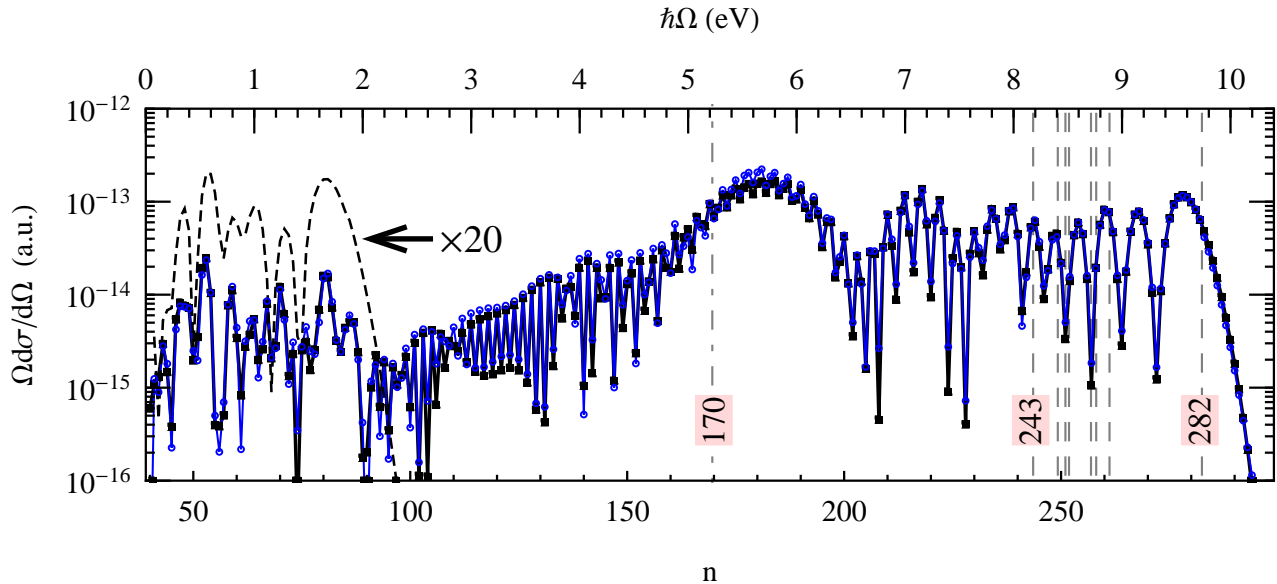}
\end{center}
\caption{(Color online) Spectral density of laser-assisted $e$-H
BrS as a function of the number $n$ of laser photons absorbed for
the same parameters of the initial and final electrons and the
laser field as in figure~\ref{Omega:t}. Thick (black) line with
squares~-- exact TDER results; thin (blue) line with circles~--
results for nonresonant rescattering in the low-frequency approximation; thin dashed
(black) line~-- results obtained using (\ref{tdp1}),
(\ref{resc2:approx}) and (\ref{resc2:1term});
the vertical gray lines indicate extremal values of
$\ov{\Omega}_{\nu}$: dashed lines~-- local maxima of
$\mathcal{E}_2(t)$, dot-dashed line~-- a local minimum of
$\mathcal{E}_2(t)$.}
 \label{Spectr}
\end{figure*}

The LABrS spectral density $\Omega d\sigma/d\Omega$ [cf.~(\ref{CS})] as a function of the emitted
photon energy $\hbar\Omega$ (or the number $n$ of laser photons exchanged) is presented in
figure~\ref{Spectr}. One sees that the LABrS spectrum clearly exhibits a plateau structure: $\Omega
d\sigma/d\Omega$ is an oscillating function for a broad interval of frequencies
$\Omega<\Omega_{\max}$ (where $\Omega_{\max}\approx 9.6$~eV defines the plateau cutoff), while
beyond the cutoff (for $\Omega>\Omega_{\max}$) the spectral density decays rapidly.
Figure~\ref{Spectr} shows the good agreement of our analytic low-frequency results [i.e., the
results~(\ref{tdp1}) and~(\ref{tdp2}) and their further modifications in section~3.5] with the
exact numerical TDER results (cf. the discussion of equations~(\ref{hatdip}) and~(\ref{Scatt-Scatt}) in section~\ref{GenResults}).

The oscillation patterns of the LABrS cross section originate from
interference of partial amplitudes given by different summands in
(\ref{tdp1}) and~(\ref{tdp2}). The contributions of the ``direct''
term $\vk{d}_n^{\dir}$ in figure~\ref{Spectr} are negligible. The
rescattering term $\vk{d}_n^{\rescI}$ (which corresponds to
spontaneous emission upon electron-atom rescattering) plays a role in the
low-energy part of the spectrum, where this term interferes with
$\vk{d}_n^{\rescII}$ (cf.~the dashed line in figure~\ref{Spectr}, which for visibility gives the contribution of the amplitude
$\vk{d}_n^{\rescI}$ multiplied by a factor of 20). Therefore, the plateau-like behavior of the
LABrS spectrum is mainly described by the amplitude
$\vk{d}_n^{\rescII}$, which corresponds to the rescattering
scenario II, with spontaneous photon emission at the first
electron-atom collision. There are only two real saddle points
$\mcal{P}_{1}$ and $\mcal{P}_{2}$ (i.e., only two terms in the sum
for $\vk{d}_n^{\rescII}$) in the plateau cutoff region. These two
saddle points coalesce [$\mcal{P}_{1} \to\ov{\mcal{P}}_{1}$,
$\mcal{P}_{2} \to\ov{\mcal{P}}_{1}$] for $\Omega=\ov{\Omega}_1
\equiv\Omega_{\max}=9.74$~eV or $n_{\max}=282$
[cf.~figures~\ref{traj}(a) and~\ref{Omega:t}]. Therefore, the
LABrS amplitude near the rescattering plateau cutoff is well
approximated by the result~(\ref{resc2:approx}),
 $\vk{d}_n^{\rescII}\approx \ov{\vk{d}}_{n,1}^{\rescII}$.
The few oscillation maxima of the spectral density closest to the
cutoff are described by the Airy function and its derivative
[involved in the propagation factors $\mcal{W'}_{k_{1,2}}^{(\nu)}$
in~(\ref{resc2:approx})], which oscillate for negative arguments,
$\Omega<\ov{\Omega}_1$ [cf.~(\ref{Airy:arg})]. For smaller
energies $\hbar\Omega$, the number of saddle points one must take
into account increases (cf.~the LABrS spectrum region
$\hbar\Omega\lesssim9$~eV in figure~\ref{Spectr}, in which the
vertical dashed lines indicate the appearance of pairs of real
saddle points), while the accuracy of the
result~(\ref{resc2:approx}) worsens. As mentioned above, the
contributions of partial amplitudes with different travel times
$\Delta t_k=|t_k'-t_k|$ decreases with increasing $\Delta t_k$. We
do not take into account saddle points ($t_k,t'_k$) that describe
long closed trajectories having traveling times $\Delta t_k>4T$
since the main contributions are given by the relatively few
trajectories with $\Delta t_k<T$. In figure~\ref{Spectr}, the
appearance of two real saddle points for $n<243$ (these solutions
correspond to closed trajectories with short travel times $\sim
0.6T$, cf.~figure~\ref{Omega:t}) leads to strong interference of
the corresponding partial amplitudes with other terms
in~(\ref{tdp2}) and, hence, to the appearance of high-frequency
oscillations in LABrS spectra.

For 5.2~eV~$\lesssim\hbar\Omega\lesssim 6$~eV, the LABrS spectral density in
figure~\ref{Spectr} exhibits a pronounced enhancement (similar to an
interference maximum in the cutoff region). This enhancement is
due to the interference of two coalescing trajectories in the
vicinity of the local minimum 
of the function $\mathcal{E}_2(t)$, $\hbar\ov{\Omega}_{\nu}=5.2$
eV (cf.~figure~\ref{Omega:t}).
The relative suppression of the LABrS cross section occurs for $\hbar\Omega \lesssim 5$~eV, because
the shortest real trajectories disappear in this region (cf.~the black curves in figure~\ref{Omega:t}).

\begin{figure}
\begin{center}
\includegraphics[width=1.\linewidth]{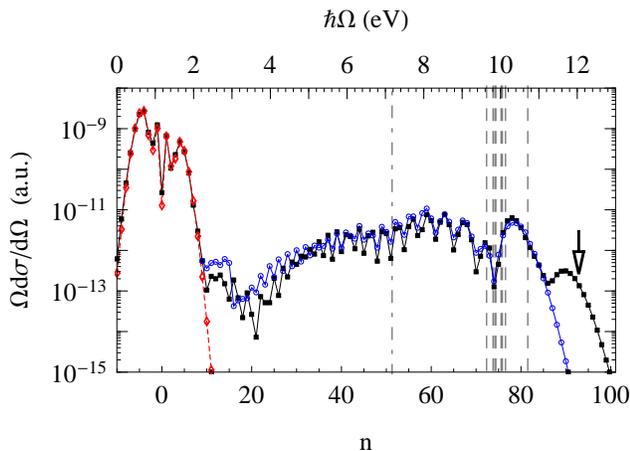}
\end{center}
\caption{(Color online) The same as in figure~\ref{Spectr} but for LABrS in a CO$_2$ laser field with
$\hbar\omega=0.117$ eV, $I=6.4\times 10^{10}$~W/cm$^2$, $E_i=4.84$~eV and $E_f=3.67$~eV. Dashed
(red) line with diamonds is the `direct' LABrS result~(\ref{DIPZR}).
The arrow indicates the cutoff of the resonant rescattering
plateau in accordance with~(\ref{LARA:cutoff}).
 }
 \label{Spectr:CO2}
\end{figure}

In figure~\ref{Spectr:CO2} we present the $e$-H LABrS spectrum for the case of a CO$_2$ laser field
with $\hbar\omega=0.117$~eV and intensity $I=6.4\times 10^{10}$~W/cm$^2$. The low-energy part of
the spectrum (for $\hbar\Omega<2$~eV) represents a plateau-like structure and is well described by
the ``direct'' part $\vk{d}_n^{\dir}$ of the LABrS amplitude.  The second (high-energy) plateau for
2.5~eV$<\hbar\Omega<11$~eV is described by the rescattering amplitude $\vk{d}_n^{\rescII}$, while
the contribution of the amplitude $\vk{d}_n^{\rescI}$ is negligible and is masked by the
``direct'' LABrS plateau. The averaged value of the spectral density along the rescattering plateau
is about 2~--~3 orders of magnitude smaller than for the ``direct'' plateau, in agreement with the
fact that the relative magnitude of the amplitudes $\vk{d}_n^{\resc}$ and $\vk{d}_n^{\dir}$ is
governed by the parameter $\mcal{A}/\alpha_0$, which is the ratio of the characteristic field-free
scattering amplitude $\mathcal{A}$ to the quiver radius, $\alpha_0=|e| F/(m\omega^2)$, of a free
electron in the laser field. Indeed, it follows from (\ref{tdp1}) and~(\ref{tdp2}) that the
rescattering amplitude $\vk{d}_n^{\resc}$ contains an additional factor $\sim\mathcal{A}/\alpha_0$
[the propagation factors $\mathcal{W}_k$ and $\mathcal{W}_k'$ are proportional to $\alpha_0^{-1}$,
cf.~(\ref{propag})] in comparison  with $\vk{d}_n^{\dir}$ (cf.~also the low-frequency analysis of
the LAES process in~\cite{LAES2013}). For low-energy $e$-H scattering within the effective range
theory, $\mcal{A}\approx a_0=6.2$~a.u., while $\alpha_0=73$~a.u. for the parameters of the spectrum in
figure~\ref{Spectr:CO2}. The strong enhancement of the spectral density beyond the rescattering
plateau cutoff (for $\hbar\Omega>11$~eV) is related to resonant LABrS due to radiative
recombination into an intermediate quasibound state of the H$^{-}$ ion~\cite{ResBrs}. As shown
in~\cite{ResBrs} (cf. also~\cite{kuch, jaron}), the cutoff of the LARA/LARR process is determined
by the relation:
\begin{equation}
\label{LARA:cutoff} \hbar\Omega^{(r)}=\frac{1}{2m}\left(\vk{p}_i+
\m{sgn}(\vk{p}_i\cdot \vk{e}_z)\vk{e}_z \frac{|e|  F}{\omega}
\right)^2 + |E_0|,
\end{equation}
where $E_0$ is the energy of the bound state. For the parameters applicable to figure~\ref{Spectr:CO2} and
$E_0=-0.755$~eV (the ground state energy of H$^{-}$), equation~(\ref{LARA:cutoff}) gives
$\hbar\Omega^{(r)} = 12.04$~eV, which coincides with the cutoff of the ``extended'' plateau in
figure~\ref{Spectr:CO2}. The disagreement of the nonresonant low-frequency results with the exact
TDER theory results for energies 2.5~eV$<\hbar\Omega<4$~eV (where the LABrS spectrum exhibits a suppression similar
to the suppression in figure~\ref{Spectr} for 2~eV$<\hbar\Omega<4$~eV) is also caused by the
previously discussed resonant channel.

\subsection{Estimate of the LABrS cross section for electron scattering from a Coulomb potential}

\begin{figure}
\begin{center}
\includegraphics[width=1.\linewidth]{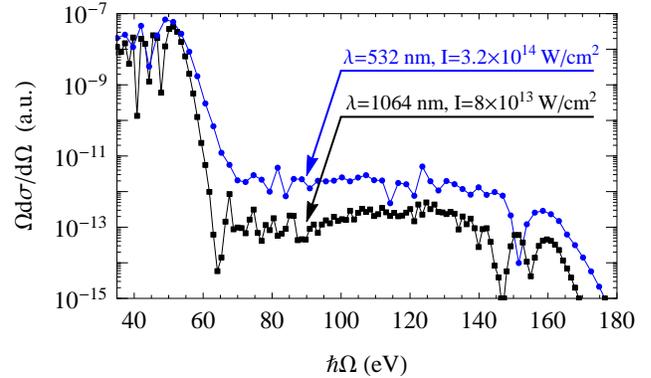}
\end{center}
\caption{(Color online) Spectral density of electron-proton BrS in laser fields with two different
wave lengths $\lambda$ and intensities $I$, as shown in the figure. Incident and final electron parameters:
$\vk{p}_{i}\|\vk{p}_{f}\|\vk{e}_z$, $E_{i}=80$~eV and $E_{f}=45$~eV.
\label{Spectr:C} }
\end{figure}

Since the analytic results~(\ref{DIPZR}), (\ref{tdp1})
and~(\ref{tdp2}) contain field-free atomic factors evaluated at
laser-modified instantaneous momenta, they allow one to extend
their applicability beyond the assumptions of the TDER theory.
This extension is very straightforward and consists in the
replacement of the atomic factors $\mcal{A}$ and $\vk{d}_0$
obtained within the TDER approach by their counterparts for a real
atomic potential. We emphasize that the propagation factors
$\mathcal{W}_k$ and $\mathcal{W}_k'$ (that describe key aspects of
rescattering in the LABrS process) do not involve any information
about electron-atom interactions and are general for any atomic
target. In figure~\ref{Spectr:C} we present LABrS spectra for the
case of the electron-Coulomb interaction, i.e. for electron-proton
BrS. The field-free quantities for this case are the
electron-Coulomb scattering amplitude $\mathcal{A}$, obtained
from~\cite{Land3}, and the BrS dipole moment $\mathbf{d}_0$,
obtained from~\cite{Land4}. In figure~\ref{Spectr:C} we present
results for two different sets of laser field parameters: (1)
$\lambda=1064$~nm, $I=8\times10^{13}$~W/cm$^2$ and (2)
$\lambda=532$~nm, $I=3.2\times10^{14}$~W/cm$^2$, such that the
ratio $F/\omega$ is the same, while the quiver radius $\alpha_0$
changes by a factor of 2. We note that both parameters $\omega
p_i/(|e|F)=2.2$ and  $\omega p_f/(|e|F)=1.6$ are greater than one;
if they were less than one the electron momenta $P_{i,f}(t'_k)$ in
the Coulomb scattering amplitude might vanish, so that the
results~(\ref{tdp1}) and~(\ref{tdp2}) would become inapplicable.
For the chosen parameters, the saddle points $\mathcal{P}_k$ (and
field-free amplitudes) are the same in both cases, while the
propagation factors differ and cause different oscillation
patterns in the LABrS spectra. As is seen in
figure~\ref{Spectr:C}, for the shorter wavelength (and thus for
the smaller quiver radius, $\alpha_0\propto\lambda^{2}\sqrt{I}$)
the spectral density is approximately 4 times higher.

\section{Concluding Summary}

In this paper, we have developed an analytic description of
the LABrS process taking into account the rescattering effects of
the active electron on the atomic target. These effects are
responsible for the occurrence of the high-energy plateau in the
dependence of the LABrS spectral density on the emitted photon
energy (or on the number of laser photons exchanged). The key results
of our analytic approach are the expressions~(\ref{tdp1})
and~(\ref{tdp2}), which present the LABrS dipole moment (the LABrS
amplitude) in the nonresonant low-frequency approximation.  These
results have a transparent interpretation in terms of the
rescattering scenario. Moreover, these results describe two
possible realizations of this scenario, scenario I and scenario
II. The result~(\ref{tdp1}) for $\vk{d}_{n}^{\rescI}$ represents
an interference of partial amplitudes, each related to
the pair of times $t_k$ and $t_k'$.  It describes the following
three-step picture (scenario I): (i) the electron elastically
scatters from the atom at the moment $t_k'$, (ii) the laser field
returns the electron back to the atom at the moment $t_k$
($t_k>t_k'$), where (iii) the electron rescatters with
sponataneous photon emission. Similarly, the result~(\ref{tdp2})
for $\vk{d}_{n}^{\rescII}$ describes the scenario II: (i) the BrS
process happens at the first electron-atom collision at the time
$t_k$, (ii) the laser field returns the electron back to the atom
at the moment $t_k'$ ($t_k'>t_k$), followed by (iii) the elastic
electron scattering (the rescattering event). The pair of times
$(t_k,t_k')$ corresponds to some closed trajectory of the electron's
motion in the laser field between the events of the first and
second electron-atom collisions. We have found that the scenario II
with spontaneous photon emission during the first collision is
allowed for an arbitrary incident electron energy $E_i$. Numerical
analysis shows that this scenario is significant for such values
of the final electron energy $E_f$ that $E_f<10u_p$. Another
situation is realized for the scenario I, with photon emission
during the second collision (rescattering). For this case, the
dipole moment $\vk{d}_{n}^{\rescI}$ can contribute to the LABrS
amplitude for $E_i<10u_p$, and exceeds the term
$\vk{d}_{n}^{\rescII}$ only for $E_i\lesssim 0.2u_p$ (for the case
of the parallel geometry $\vp_i\|\vp_f\|\ve_z$). In comparison
with the direct LABrS process, the rescattering effects
significantly extend the maximum energy of the emitted photon (up
to the rescattering plateau cutoff), while the averaged value of
the spectral density is about 2~--~3 orders of magnitude smaller
than for the direct process. Finally, the clear physical meaning
of the key factors involved in the TDER results~(\ref{tdp1})
and~(\ref{tdp2}) allow us to generalize those factors to the case of real
atoms or ions. Such generalization has been made for
electron-proton LABrS, followed by the numerical evaluation of
the LABrS spectral density for this case.

\ack

This work was supported in part by RFBR Grant nos~14-02-31412 young$_{-}$a and 13-02-00420, by NSF
Grant no~PHYS-1208059 and by the Ministry of Education and Science of the Russian Federation
(project no 1019). ANZ acknowledges the support of the `Dynasty' Foundation.

\appendix

\section*{Appendix. Derivation of results (\ref{tld+}) and (\ref{tld-}) for $\tld$}

\setcounter{section}{1}

Taking into account~(\ref{Phi+}) and~(\ref{Phi-}), we rewrite the definition~(\ref{Scatt-Scatt})
for $\tld$ in explicit form:
\begin{eqnarray}\nonumber
&& {\tld} =
  e\left(\frac{2\pi \hbar^2}{m \kappa}\right)^2
  \frac{1}{T}\int\limits_{0}^{T}dt
\int\limits_{-\infty}^{t} dt'  \int\limits_{t}^{\infty} dt'' \\
\nonumber &\times& e^{i n \omega t+i[\epsilon_{i}
(t-t')+\epsilon_f (t''-t)]/\hbar}
f_{\vk{p}_{i}}(t')f_{-\vk{p}_f}(-t'') \\
&\times& \int d\vk{r}\,G^{(-)*}(\vr,t;0,t'') \, \vk{r}\,
G^{(+)}(\vr,t;0,t') .\hspace{5mm} \label{app:d3}
\end{eqnarray}
For the spatial integral in (\ref{app:d3}), we use the following relation (cf.~the Appendix in
\cite{Descr_gen}):
\begin{eqnarray*}
&&\int d\vk{r}\, G^{(+)}(\vr,t;0,t')\, \vk{r}\, G^{(-)*}(\vr,t;0,t'') \\
&&=
\frac{i}{\hbar}G^{(+)}(0,t'';0,t')\boldsymbol{\mathcal{R}}(t;t',t''),
\end{eqnarray*}
where
\begin{eqnarray*}
\boldsymbol{\mathcal{R}}(t;t',t'')&=&\frac{e}{m\omega^2(t'-t'')}
\Big\{(t-t'')\big[\vk{F}(t)-\vk{F}(t')\big] \\
&-& (t-t')\big[\vk{F}(t)-\vk{F}(t'')\big]\Big\}.
\end{eqnarray*}
In order to treat the three-dimensional integral over $t''$, $t'$,
$t$ in (\ref{app:d3}), we introduce new variables: $\xi =
(t''-t')/2$,  $\zeta = t-(t'+t'')/2$, $\tl{t}=t-\zeta$
($0\leqslant\xi<\infty$, $- \xi \leqslant \zeta \leqslant \xi$).
Integration over $\zeta$ leads to the result:
\begin{eqnarray}
&& {\tld}={\tld}^{(+)}+{\tld}^{(-)}, \nonumber \\
&& {\tld}^{(\pm)}=\mp\frac{e^2}{ 2 \omega^2\Omega}\sqrt{\frac{ \pi
i \hbar}{ m^3}}\frac{1}{T}\int_0^Td\tl{t}
\int\limits_{0}^{\infty} \frac{d\xi}{\xi^{3/2}} e^{\pm i\xi \Omega} \nonumber \\
&& \times e^{i n\omega \tl{t}
+(i/\hbar)\left[(\epsilon_{i}+\epsilon_{f})
\xi+S\left(\tl{t}+\xi,\tl{t}-\xi\right)\right]}\nonumber
\\
&&\times f_{\vk{p}_{i}}\left(\tl{t}-\xi\right)
f^{}_{-\vk{p}_{f}}\left(-\tl{t}-\xi\right)\Bigg\{
\frac{\vk{F}\left(\tl{t}-\xi\right)-
   \vk{F}\left(\tl{t}+\xi\right)}{\Omega\xi}\nonumber
\\
&& +2 i \vk{F}\left(\tl{t}\pm \xi\right)-i\vk{e}_z F\sum_{s=\pm 1}
\frac{\Omega  e^{i s\omega (\tl{t}\pm \xi)}}{\Omega+s \omega}
\Bigg\}.
\label{two_part_dpm}
\end{eqnarray}
Applying the variable replacement $\tl{t}=t-\xi $ to $\tld^{(+)}$ in~(\ref{two_part_dpm}) followed
by the replacement $t'=t-2\xi$, we obtain  for ${\tld}^{(+)}$ the expression~(\ref{tld+}).
Similarly, applying the variable replacement $\tl{t}=t+\xi$ to $\tld^{(-)}$ in (\ref{two_part_dpm})
followed by the replacement $t'=t+2\xi$, we obtain the result~(\ref{tld-}) for ${\tld}^{(-)}$.

\end{document}